\documentclass[a4paper,11pt]{article}

\usepackage{jheppub}

\usepackage[T1]{fontenc} % if needed

\title{\boldmath Studying gaugino masses in supersymmetric model at future 100 TeV $pp$ collider}

\author[a,b]{Shoji Asai}
\author[a]{So Chigusa}
\author[c]{Toshiaki Kaji}
\author[a]{Takeo Moroi}
\author[a]{Masahiko Saito}
\author[b]{Ryu Sawada}
\author[b]{Junichi Tanaka}
\author[b]{Koji Terashi}
\author[a]{Kenta Uno}

\affiliation[a]{Department of Physics, The University of Tokyo,\\Hongo 7-3-1, Bunkyo-ku, Tokyo 113-0033, Japan}
\affiliation[b]{International Center for Elementary Particle Physics, The University of Tokyo,\\Hongo 7-3-1, Bunkyo-ku, Tokyo 113-0033, Japan}
\affiliation[c]{Waseda University,\\Ookubo 3-4-1, Shinjuku-ku, Tokyo 169-8555, Japan}

\emailAdd{shoji.asai@cern.ch}
\emailAdd{chigusa@hep-th.phys.s.u-tokyo.ac.jp}
\emailAdd{toshiaki.kaji@cern.ch}
\emailAdd{moroi@phys.s.u-tokyo.ac.jp}
\emailAdd{masahiko.saito@cern.ch}
\emailAdd{ryu.sawada@cern.ch}
\emailAdd{junichi.tanaka@cern.ch}
\emailAdd{koji.terashi@cern.ch}
\emailAdd{kenta.uno@cern.ch}

\usepackage{subcaption}

\abstract{

   We discuss prospects of studying supersymmetric model at future $pp$
  circular collider (FCC) with its centre-of-mass energy of $\sim 100\
  {\rm TeV}$.  We pay particular attention to the model in which Wino
  is lighter than other supersymmetric particles and all the gauginos
  are within the kinematical reach of the FCC, which is the case in a
  large class of so-called pure gravity mediation model based on
  anomaly mediated supersymmetry breaking.  In such a class of model,
  charged Wino becomes long-lived with its decay length of $\sim 6\
  {\rm cm}$, and the charged Wino tracks may be identified in
  particular by the inner pixel detector; the charged Wino tracks can
  be used not only for the discrimination of standard model
  backgrounds but also for the event reconstructions.  We show that
  precise determinations of the Bino, Wino, and gluino masses are
  possible at the FCC.  For such measurements, information about the
  charged Wino tracks, including the one about the velocity of the
  charged Wino using the time of the hit at the pixel detector, is
  crucial.  With the measurements of the gaugino masses in the pure
  gravity mediation model, we have an access to more fundamental
  parameters like the gravitino mass.

}


\begin{document}

\maketitle
\flushbottom

\newcommand{\lsim}{\stackrel{<}{_\sim}}
\newcommand{\gsim}{\stackrel{>}{_\sim}}
\newcommand{\hyphen}{\,\mathchar`-\mathchar`-\,}

\section{Introduction}
\label{sec:intro}

There are strong motivations to consider physics beyond the standard
model (BSM) at TeV scale.  One is dark matter that cannot be explained
in the framework of the standard model; particularly, the scenario of
thermal relic dark matter requires the annihilation cross section of the
dark matter particle to be $\sim 1\ {\rm pb}$, which is the typical
cross section for particles with the mass scale $\sim 0.1$--$1\ {\rm TeV}$.
In addition, a large hierarchy between the electroweak scale and the
Planck scale, i.e.~the scale of the gravity, motivates us to consider
some mechanism to relax the quadratic divergence in the Higgs mass
parameter in the standard model.

High-energy-collider experiments are crucial for the study of BSM
physics.  LHC experiments have been searching for the BSM
particles as an energy-frontier, but no direct sign has
been discovered yet.  Although LHC experiments will keep on
exploring BSM particles at higher mass scale, BSM particles may be
just above the discovery reach of the LHC.  In this case, a collider with
higher energy than the LHC is needed to discover and study the BSM
physics.  Among possibilities, we consider $pp$ option of Future
Circular Collider (FCC) with the centre-of-mass energy of $\sim 100\
{\rm TeV}$ \cite{Mangano:2016jyj, Contino:2016spe, Golling:2016gvc}
because it can drastically push up the energy frontier of high-energy
experiments.  Because the FCC is considered to be one of promising
collider experiments, the understanding of the physics potential of
the FCC is of high importance.

Here, we discuss the prospect of the FCC in the study of
supersymmetric (SUSY) model.  We concentrate on the so-called pure
gravity mediation (PGM) model of SUSY breaking \cite{Ibe:2006de,
  Ibe:2011aa, ArkaniHamed:2012gw}, which is strongly motivated from
theoretical and phenomenological points of view.  As we will discuss
in the next section, in the PGM model, scalar particles in the minimal SUSY
standard model (MSSM) have masses of $O(100)\ {\rm TeV}$ and may be
too heavy to be studied at the FCC.  However, gaugino masses are
generated by the anomaly mediation \cite{Randall:1998uk,
  Giudice:1998xp} and are loop suppressed relative to the scalar
masses.  Typically, the gauginos are at the TeV scale and are within
the reach of the FCC.  Such a spectrum is easily obtained by a simple
set up of the SUSY breaking sector.  In addition, the PGM model has
several advantages in cosmology.  In particular,
a thermally produced lightest superparticle (LSP) can be a viable
candidate of dark matter under the assumption of $R$-parity
conservation \cite{ArkaniHamed:2006mb}.

One important feature of the model of our interest, the PGM model, is
that the Winos, which are the gauginos for $SU(2)_L$ gauge group, are
likely to be lighter than other superparticles.  In particular, due to
the radiative correction, the neutral Wino $\tilde{W}^0$ is slightly
lighter than charged ones $\tilde{W}^\pm$ \cite{Cheng:1998hc,
  Feng:1999fu} and hence is the LSP.  With such a mass spectrum,
charged Wino has a relatively long lifetime and, once produced at the
FCC, it may fly a macroscopic distance of $\sim 10\ {\rm cm}$ before
its decay.  The track of such a ``long-lived'' charged Wino may be
identified by the inner pixel detector of the FCC, which provides a
very characteristic and unique signal of the PGM model.  The
characteristic signal is used in searches for the Wino in LHC
experiments and exclusion limits of Wino masses are set by the
ATLAS~\cite{ATLAS:2018} and CMS~\cite{CMS:2018} experiments.  The
discovery reaches of this type of searches at the FCC have been
studied in, for example, ref.~\cite{Low:2014cba,Cirelli:2014dsa}.  The
charged Wino tracks expected in the PGM model can be used not only for
the discovery but also for the study of the properties of SUSY
particles.

In this paper, we discuss the prospect of the study of SUSY particles at
the FCC, paying particular attention to the PGM model of SUSY breaking.
We mainly focus on a parameter region where all the gauginos, i.e, Bino,
Winos, and gluino, are within the reach of the FCC, assuming that its
centre-of-mass energy is $100\ {\rm TeV}$.  We propose procedures to
measure Wino, Bino, and gluino masses at the FCC, using the pair
production process of gluinos, $pp\rightarrow\tilde{g}\tilde{g}$, and
the pair production process of charged Winos associated with hard jets,
$pp\rightarrow\tilde{W}^{+}\tilde{W}^{-}+\mathrm{jets}$.  We show that
information about the charged Wino tracks from the inner pixel detector
is crucial not only for the background reduction but also for the
reconstruction of SUSY events.  We also point out that, for the gaugino
mass determination, it is important to precisely determine the velocity
of charged Wino, which is expected to be possible by using the time
information available from the pixel detector.

The organization of this paper is as follows.  In section
\ref{sec:model}, we introduce the PGM model and introduce several
Sample Points for our Monte Carlo (MC) analysis.  In section
\ref{sec:susyevents}, we explain basic features of the SUSY events at
the FCC.  The results of the MC analysis for the gaugino mass
determinations are shown in section \ref{sec:gauginomass}.
Implications of the gaugino mass determinations for the understanding
of the underlying theory are discussed in section
\ref{sec:implications}.  Section \ref{sec:conclusions} is devoted to
conclusions and discussions.

\section{Model}
\label{sec:model}
\setcounter{equation}{0}

Let us first introduce the SUSY model we consider.  Among various
possibilities, we concentrate on the so-called pure gravity mediation
(PGM) model of SUSY breaking \cite{Ibe:2006de, Ibe:2011aa,
ArkaniHamed:2012gw} based on anomaly mediation \cite{Randall:1998uk,
Giudice:1998xp}.  In this model, the sfermion and Higgsino masses are
generated by (tree-level) Planck-suppressed operators and are of the
order of the gravitino mass $m_{3/2}$.  On the contrary, assuming that
there is no singlet field in the SUSY breaking hidden sector so that
tree-level gaugino masses in the MSSM sector are forbidden, the gaugino
masses are obtained from the effect of anomaly mediation.  In this model,
the gaugino masses are one-loop suppressed compared to the gravitino mass.
Then, the gauginos are relatively light and are the primary targets of
collider experiments.

In the PGM model, the SUSY breaking gaugino mass parameters for
$U(1)_Y$, $SU(2)_L$, and $SU(3)_C$ gauge interactions, denoted as
$M_1$, $M_2$, and $M_3$, respectively, are given as
\cite{Giudice:1998xp, Gherghetta:1999sw}
\begin{align}
  M_1 (M_{\rm S})= &\,
  \frac{g_1^{2} (M_{\rm S})}{16 \pi^{2}} \left( 11 m_{3/2} + L \right),
  \label{m1}
  \\
  M_2 (M_{\rm S}) = &\,
  \frac{g_2^{2} (M_{\rm S})}{16 \pi^{2}} \left( m_{3/2} + L \right),
  \label{m2}
  \\
  M_3 (M_{\rm S}) = &\,
  \frac{g_3^{2} (M_{\rm S})}{16 \pi^{2}} \left( -3 m_{3/2} \right),
  \label{m3}
\end{align}
where $g_1$, $g_2$, and $g_3$ are gauge-coupling constants of
$U(1)_Y$, $SU(2)_L$, and $SU(3)_C$ gauge interactions, respectively.
Here, $L$ denotes the effect of the threshold correction due to the
Higgs and Higgsino loop and is given by
\begin{align}
  L \equiv \mu \sin2\beta 
  \frac{m_{A}^{2}}{|\mu|^{2}-m_{A}^{2}} 
  \ln \frac{|\mu|^{2}}{m_{A}^{2}},
  \label{eq:L_def}
\end{align}
where $\mu$ and $m_A$ are the SUSY invariant Higgs mass and the
pseudo-scalar Higgs mass, respectively, while $\beta$ being the angle
parametrizing the relative sizes of vacuum expectation values of up- and
down-type Higgses.  In addition, $M_{\rm S}$ is the mass scale of
sfermions and Higgsinos, and hence $M_{\rm S}\sim m_{3/2}$.  Here and
hereafter, we take the convention such that $m_{3/2}$ is real and
positive.  We use the $\overline{\mathrm{MS}}$ scheme for the
calculation of gauge coupling constants and mass parameters.

In general, $M_1$, $M_2$, and $M_3$ are complex and depend on the
renormalization scale.  The gaugino masses (i.e.~mass eigenvalues) are
determined from $M_1$, $M_2$, and $M_3$ at the renormalization scale
close to the mass eigenvalues.  The mass eigenvalues of Bino, Wino, and
gluino, which are real and positive, are denoted as $m_{\tilde{B}}$,
$m_{\tilde{W}}$, and $m_{\tilde{g}}$, respectively.\footnote
{Because the Higgsino mass is much larger than the gaugino masses, the
  mixings between gauginos and Higgsinos are not important in the
  present case, and we regard Bino and Wino as mass eigenstates.  In
  addition, as we will discuss, the masses of charged and neutral
  Winos are slightly different but their difference is much smaller
  than the expected accuracy of the Wino mass determination at the
  FCC.  Thus, we neglect the mass difference unless it is important.}
We take into account the one-loop threshold correction to the gluino
mass~\cite{Giudice:2004tc}
\begin{align}
 m_{\tilde{g}} = | M_3 (M_\mathrm{G}) | \left[ 1 + \frac{g_3^2}{16\pi^2}
 \left( 12 + 9 \ln \frac{M_\mathrm{G}^2}{|M_3|^2} \right) \right],
\end{align}
where $M_\mathrm{G}$ is the mass scale of gauginos, which we determine
by solving $M_\mathrm{G} = |M_3(M_\mathrm{G})|$ in our analysis.  In
addition, we neglect the small threshold corrections to Bino and Wino at
the gaugino mass scale and just take $m_{\tilde{B}} =
|M_1(M_\mathrm{G})|$ and $m_{\tilde{W}} = |M_2(M_\mathrm{G})|$.  When
$m_{3/2}$ and $L$ are of the same order of magnitude, Wino becomes
lighter than Bino and gluino.

In our analysis, we consider the case where the gravitino mass is of
$O(100)\ {\rm TeV}$ (and hence the sfermion and Higgsino masses are),
while the gaugino masses are of $O(1)\ {\rm TeV}$.  Such a mass
spectrum is phenomenologically viable and well-motivated:
\begin{itemize}
\item The observed value of the Higgs mass (i.e.~$m_h\simeq 125\ {\rm
    GeV}$ \cite{Tanabashi:2018oca}) can be realized due to radiative
  corrections \cite{Okada:1990vk, Okada:1990gg, Ellis:1990nz,
    Ellis:1991zd, Haber:1990aw}.
\item Neutral Wino is a viable candidate of dark matter.  In
  particular, if $m_{\tilde{W}}\simeq 2.9\ {\rm TeV}$, thermal relic
  density of Wino becomes equal to the present dark matter density
  \cite{Hisano:2006nn}.  In addition, even if the Wino mass is smaller
  than $\sim 2.9\ {\rm TeV}$, Wino dark matter is possible with
  non-thermal production \cite{Giudice:1998xp, Moroi:1999zb}.
\item With the large gravitino mass of $O(100)\ {\rm TeV}$, a dangerous
  cosmological gravitino problem can be avoided even if the reheating
  temperature after inflation is as high as $O(10^9)\ {\rm GeV}$
  \cite{Kawasaki:2017bqm}.  A simple scenario of leptogenesis
  \cite{Fukugita:1986hr} works with such reheating temperature
  \cite{Buchmuller:2004nz, Giudice:2003jh}.
\item Even if the sfermion and Higgsino masses are as high as $O(100)\
  {\rm TeV}$, the gauge coupling constants still meet at $\sim
  10^{16}\ {\rm GeV}$, and the SUSY grand unified theory, which is one
  of the strong motivations to consider SUSY models, is still viable
  \cite{Giudice:2004tc}.
\item Because of large sfermion masses, CP and flavour violations due
  to the loops of SUSY particles are suppressed.  This will
  significantly relax the SUSY CP and flavour problems
  \cite{Moroi:2013sfa, McKeen:2013dma}.
\end{itemize}
Although there exist motivations to adopt the mass spectrum mentioned
above, we note here that the model of our interest requires a sizeable
tuning of parameters to realize viable electroweak symmetry breaking.
This is due to the hierarchy between the mass scale of superparticles
and the electroweak scale.  We also comment, however, that the
hierarchy here is significantly reduced compared to that between the
Planck and the electroweak scales which causes the naturalness problem
in the standard model (assuming that the cut-off scale is the Planck
scale).

Because the neutralino dark matter is one of the strong motivations of
low-energy SUSY, we consider the case that the Wino becomes lighter
than the Bino and gluino so that neutral Wino becomes the candidate of
dark matter.  (Notice that, in the present framework, the Bino hardly
becomes a viable candidate of dark matter even if it is the LSP.  This
is because, with heavy sfermions and Higgsinos, its pair annihilation
cross section is too small to realize the correct relic density to be
dark matter.)  In particular, we assume that the Wino mass is $\sim
2.9\ {\rm TeV}$ so that the thermal relic density of Wino becomes
consistent with the present dark matter density.

In the following, we adopt several Sample Points for our MC analysis.  The
mass spectrum and fundamental parameters are summarized in table
\ref{table:samplept}.  As shown in table \ref{table:samplept},
$m_{\tilde{W}}<m_{\tilde{B}}<m_{\tilde{g}}$ in all Sample Points.

Assuming $R$-parity conservation, gluino decays as
\begin{align*}
  \tilde{g} \rightarrow
  \bar{q} q \tilde{B}, ~~~
  \bar{q} q^{(\prime)} \tilde{W},
\end{align*}
where $q$ and $\bar{q}$ denote the standard-model quark and
anti-quark, respectively.  The branching ratio for each decay mode
depends on the mass spectrum of squarks.  If the masses of left- and
right-handed squarks are the same, the branching ratio for the latter
process becomes larger than the former.  However, if the mass of the
right-handed squarks are lighter than the left-handed ones, they can
be comparable (or $\mathrm{Br}\, (\tilde{g} \rightarrow \bar{q} q
\tilde{B})$ may even become larger than $\mathrm{Br}\, (\tilde{g}
\rightarrow \bar{q} q^{(\prime)} \tilde{W})$).  The mass spectrum of
the squarks is strongly model-dependent; it depends on dimension-6
operators connecting (s)quark chiral multiplets and SUSY breaking
fields in the K\"ahler potential.  Because the detail of the squark
mass spectrum is currently unknown, we simply assume that
\begin{align}
  \mathrm{Br}\, (\tilde{g} \rightarrow \bar{q} q \tilde{B})
  = 
  \mathrm{Br}\, (\tilde{g} \rightarrow \bar{q} q^{(\prime)} \tilde{W})
  = 0.5,
\end{align}
and that the decay process is flavour universal.  Such branching
ratios are realized when the masses of right-handed squarks are
smaller than those of left-handed ones by a factor of a few.  We note
here that one of the motivations of the assumptions mentioned above are
to demonstrate the possibility of the Bino mass determination as we
explain in the following.  The dominant decay modes of Bino are given
by
\begin{align*}
  \tilde{B} \rightarrow
  \tilde{W}^\pm W^\mp, ~~~
  \tilde{W}^0 h,
\end{align*}
and, when $|\mu|\gg |M_{1,2}|\gg m_{W,h}$ (with $m_{W}$ and $m_h$
being the $W$-boson mass and Higgs mass, respectively), the decay
rates are approximately given by\footnote
{Rigorously speaking, the ``Bino'' and ``Wino'' indicate the mass
  eigenstates which consist mostly of Bino and Wino, respectively.}
\begin{align}
 \Gamma_{\tilde{B}\to \tilde{W}^{\pm} W^{\mp}} &=
 \frac{\beta_{\tilde{W}^{\pm} W^{\mp}} \kappa^2}{8\pi}
 m_{\tilde{B}} \left( 1+\frac{m_{\tilde{W}}}{m_{\tilde{B}}} \right)^2
 \left[ 1 + \frac{2m_W^2}{(m_{\tilde{B}} + m_{\tilde{W}})^2} \right]
 \left[ 1 - \frac{m_W^2}{(m_{\tilde{B}} - m_{\tilde{W}})^2} \right],\\
 \Gamma_{\tilde{B} \to \tilde{W}^0 h} &=
 \frac{\beta_{\tilde{W}^0 h} \kappa^2}{8\pi} m_{\tilde{B}}
 \left[ \left( 1 + \frac{m_{\tilde{W}}}{m_{\tilde{B}}} \right)^2
 - \frac{m_h^2}{m_{\tilde{B}}^2} \right],
\end{align}
where
\begin{align}
 \beta_{\tilde{W}^{\pm} W^{\mp}}^2 =
 \frac{m_{\tilde{B}}^4 - 2(m_{\tilde{W}^{\pm}}^2 + m_W^2) m_{\tilde{B}}^2
 + (m_{\tilde{W}^{\pm}}^2 - m_W^2)^2}{m_{\tilde{B}}^4},
\end{align}
and $\beta_{\tilde{W}^0 h}^2$ is obtained by replacing
$m_{\tilde{W}^{\pm}} \to m_{\tilde{W}^0}$ and $m_W \to m_h$ in the
above formula.  In addition, $\kappa \equiv g_1 g_2 v \sin\beta
\cos\beta \mu^{-1}$ with $v\simeq 174\,\mathrm{GeV}$ being the vacuum
expectation value of the Higgs boson.  The masses of charged and
neutral Winos split after the electroweak symmetry breaking.  When
$|\mu|$ is much larger than the electroweak scale, which is the case
in the present framework, the mass splitting is dominantly via
radiative correction due to loop diagrams with electroweak gauge
bosons \cite{Cheng:1998hc, Feng:1999fu}, and the neutral Wino becomes
lighter than the charged one.  Based on the two-loop calculation, the
mass splitting is $\delta m_{\tilde{W}}\simeq 165\ {\rm MeV}$
\cite{Ibe:2012sx}; then charged Wino dominantly decays as
$\tilde{W}^\pm\rightarrow\tilde{W}^0\pi^\pm$ with
\begin{align}
  c \tau_{\tilde{W}^\pm} \simeq 5.75\ {\rm cm},
\end{align}
where $\tau_{\tilde{W}^\pm}$ is the proper lifetime of charged Wino while $c$
is the speed of light.

\begin{table}[t]
  \begin{center}
    \begin{tabular}{c|ccc}
      \hline\hline
      & Point 1 & Point 2 & Point 3 
      \\
      \hline
      $m_{3/2}$ [TeV]
      & $250$ & $302$ & $350$
      \\
      $L$ [TeV]
      & $800$ & $756$ & $709$
      \\
      $m_{\tilde{B}}$ [GeV] & $3660$ & $4060$ & $4470$
      \\
      $m_{\tilde{W}}$ [GeV] & $2900$ & $2900$ & $2900$
      \\
      $m_{\tilde{g}}$ [GeV] & $6000$ & $7000$ & $8000$
      \\
      $\sigma (pp\rightarrow \tilde{g}\tilde{g})$ [fb]
      & $7.9$ & $2.7$ & $1.0$
      \\
      \hline\hline
    \end{tabular}
    \caption{Fundamental parameters ($m_{3/2}$ and $L$), gaugino
      masses, and the gluino pair production cross section (for the
      centre-of-mass of $100\ {\rm TeV}$), for Sample Points 1, 2, and
      3.}
    \label{table:samplept}
  \end{center}
\end{table}

\section{SUSY events at the FCC}
\label{sec:susyevents}
\setcounter{equation}{0}

In this section, we discuss important features of SUSY events at the FCC
which are used for our analysis.  In the sample points of our choice,
gluino is within the kinematical reach of the FCC, and its pair
production process is the primary target.  Hereafter, we consider the
pair production process of gluinos, $pp\rightarrow\tilde{g}\tilde{g}$,
and the pair production process of charged Winos associated with a
high-$p_T$ jet, $pp\rightarrow\tilde{W}^{+}\tilde{W}^{-}+\mathrm{jets}$.
From these processes we extract information about the gaugino masses
as we discuss in the following.

\subsection{Background estimation}

In order to eliminate standard-model backgrounds, we use the fact that
charged Wino tracks, which are disappearing and high-$p_T$, may be
recognized by the inner pixel detector.  In the gluino pair production
events, each gluino decays down to a charged or neutral Wino.  Although
charged Wino is unstable, it has a sizeable lifetime and
its $c\tau_{\tilde{W}^\pm}$ is about $5.75\ {\rm cm}$ that
is of the same order of the distance to the pixel detector
from the interaction point.  Thus, once a charged Wino is produced, it
may hit several layers of the pixel detector before its decay into a
neutral Wino and a pion.
Charged Wino track is expected to be (i)
short (i.e.~disappearing in the tracker), and (ii) high-$p_T$; such a track is
hardly realized by a single standard-model particle.
Another source of fake Wino-like track is due to an accidental
alignment of hits of particles mainly from pile-ups and it is the dominant source
according to ref.~\cite{Saito2019}.
In addition, in the signal events (i.e.~the gluino
pair production events), the missing transverse momentum evaluated
only from jets is likely to be large because of the momenta carried
away by Winos; this is not expected in the background events that have
fake Wino-like tracks.

The gaugino mass reconstruction in our analysis is based on the events
satisfying the following requirements:
\begin{itemize}
\item {\it Requirement 1}: There is a ``long enough'' Wino-like track.
  The transverse length of the track, denoted as $L_{T,1}$, should be
  longer than $L_{T,1}^{\rm (min)}=10\ {\rm cm}$.  In addition, the
  pseudo-rapidity ($\eta$) of the track is required to be $|\eta|<1.5$.
\item {\it Requirement 2}: There is another ``long enough'' Wino-like track.
  The transverse length of the track, denoted as $L_{T,2}$ should be longer
  than $L_{T,2}^{\rm (min)}$.  As we will discuss later,
  $L_{T,2}^{\rm (min)}$ is set to be $10\ {\rm cm}$ but
  results with $L_{T,2}^{\rm (min)}=5\ {\rm cm}$ are also shown
  to see how the event selection affects on the mass determination.
  In addition, the pseudo-rapidity of the track is required to be $|\eta|<1.5$.
\item {\it Requirement 3}: The missing transverse momentum should be
  larger than $1\ {\rm TeV}$.
\end{itemize}
We use signal events in which the decay chain of each gluino contains a charged Wino.
We assume that the charged Wino with a longer (shorter) transverse
flight length in an event can be fully identified when the transverse
flight length is longer than $L_{T,1}^{\rm (min)}$ ($L_{T,2}^{\rm (min)}$).

According to ref.~\cite{Saito2019}, the probability to have one fake
disappearing track in the pseudorapidity range of $|\eta|<1.5$ is
about $2\times 10^{-5}$ per bunch crossing, requiring hits in five
pixel layers for a particular setup of the FCC experiment.
The background rate can be reduced therefore by ten orders of magnitude
by requiring both Requirement 1 and Requirement 2.
Additionally, the Requirement 3 can also reduce them as well.
As a result, standard-model backgrounds can be negligible by applying these requirements.

\subsection{Wino velocity measurement}

We assume that the information about the velocity of the charged Wino
will be available if the flight length is long enough.  It may be
possible that each pixel layer determines the time of the hit with the
resolution of $O(10)\ {\rm ps}$ by utilizing, for example, low-gain
avalanche detectors~\cite{Pellegrini2014, Lange2017}, with which the
velocity relative to the speed of light ($\beta$) of the charged Wino
could be determined with the accuracy of $O(1)\ \%$.  We estimate the
expected accuracy of the $\beta$ measurement via a simple MC analysis.
We generate straight tracks, and smear the time of hits at each
pixel-detector layers to determine the ``observed'' times of hits.
The time of the $pp$ collision (which can be identified as the time
when Winos are produced) is assumed to be accurately determined by
using associated jets.  Then, the observed $\beta$ is determined for
each track by fitting the observed times of hits by a linear function
assuming a constant $\beta$.  The accuracy is estimated as the
standard deviation of the distribution of the difference between the
observed and the generator-level velocities as a function of the
generator-level $\beta$.  Then the average accuracy for Wino tracks is
estimated by taking the weighted average using the generator-level
$\beta$ distribution in our Wino samples.  We found that, assuming the
hit-level time resolution of $20\ {\rm ps}$ ($40\ {\rm ps}$), the Wino
velocity can be determined with the accuracy of $3.0\ \%$ and $3.3\
\%$ ($6.1\ \%$ and $6.8\ \%$) for charged Winos reaching the pixel
layer located at the transverse length of $10\ {\rm cm}$ with
$\beta<0.8$ and $0.9$, respectively.  We require that the
transverse flight length should be longer than $10\ {\rm cm}$ for the
$\beta$ measurement, and assume that the resolution of the observed
Wino $\beta$ is $6\ \%$.

\subsection{MC simulation}

\begin{figure}[t]
  \centerline{\epsfxsize=0.5\textwidth\epsfbox{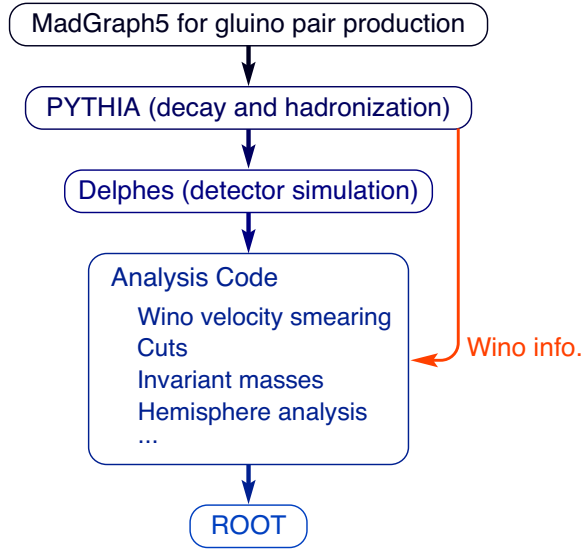}}
  \caption{\small The flowchart of our MC analysis.}
  \label{fig:flowchart}
\end{figure}

The flowchart of our MC analysis is shown in figure \ref{fig:flowchart}.
We use {\tt MadGraph5\_aMC@NLO} ({\tt v2.6.3.2}) \cite{Alwall:2011uj,
Alwall:2014hca} to generate $pp\rightarrow\tilde{g}\tilde{g}$ and
$pp\rightarrow\tilde{W}^{+}\tilde{W}^{-}+\mathrm{jets}$ events.  Then,
decay and hadronization processes are taken care of by {\tt PYTHIA8}
\cite{Sjostrand:2014zea}.  A fast detector simulation is performed by
{\tt Delphes} ({\tt v3.4.1}) \cite{deFavereau:2013fsa}; we use the card
{\tt FCChh.tcl} included in the package.  Because the velocity
measurement of charged Wino cannot be simulated by the {\tt Delphes}
package by default, the information about Winos provided by {\tt
PYTHIA8}, as well as {\tt Delphes} outputs, are passed to our original
analysis code; the analysis code smears the velocities of charged Winos
to determine observed velocities, applies kinematical cuts, reconstructs
kinematical variables (including invariant masses that will be discussed
in the following), and performs the hemisphere analysis.  In particular,
for charged Winos whose transverse flight length $L_T$ is longer than
$10\ {\rm cm}$, the observed value of the Wino velocity is determined
by\footnote
{Here and hereafter, $\beta$ is used for the observed Wino velocity;
  it should not be confused with the ratio of the Higgs VEVs.}
\begin{align}
  \beta = (1 + 6\ \% \times Z) \beta^{\rm (true)},
\end{align}
where $\beta$ and $\beta^{\rm (true)}$ are observed and true values of
the velocity, and $Z$ is the $(0,1)$ Gaussian random variable.  An
accurate velocity determination may be difficult for charged Winos
with a short transverse flight length, for example, 
$L_T<10\ {\rm cm}$; thus Winos with $L_T<10\ {\rm cm}$ are not used
in the mass determination.
The fittings of invariant-mass distributions are performed by {\tt ROOT}
(\texttt{v6.14}) \cite{Brun:1997pa}.

\section{Gaugino mass determination}
\label{sec:gauginomass}
\setcounter{equation}{0}

Now we are at the position of discussing how and how well we can
determine the gaugino masses.  In order to demonstrate that the gaugino
mass determinations are possible, we generate
$pp\rightarrow\tilde{g}\tilde{g}$ and
$pp\rightarrow\tilde{W}^{+}\tilde{W}^{-}+\mathrm{jets}$ events.  Then,
we study the reconstructions of gaugino masses using data sets which
correspond to multiple experiments with a fixed integrated luminosity ${\cal
L}$ for each experiment.  As for $pp\rightarrow \tilde{g}\tilde{g}$ process, we
take ${\cal L}=10\ {\rm ab}^{-1}$ for Sample Points 1 and 2, while
${\cal L}=30\ {\rm ab}^{-1}$ for Sample Point 3.  For $pp\rightarrow
\tilde{W}^{+}\tilde{W}^{-}+\mathrm{jets}$ process, we take ${\cal L}=30\
{\rm ab}^{-1}$.

\subsection{Wino mass determination}

We first consider the measurement of the Wino mass.  Using charged
Winos identified as short high-$p_T$ tracks with velocity information,
we may determine Wino mass if the momentum of the Winos are
known. Even though the curvature of the track depends on the Wino
momentum, most of the Winos are so high-$p_T$ that precise
determination of the curvature is expected to be unlikely.

Instead of using the curvature information, we use the conservation of
the transverse momentum to determine the Wino momenta.  Because we use
events with two charged Wino tracks, the directions of the Wino
momenta (which we denote $\vec{n}_{\tilde{W}_1^\pm}$ and
$\vec{n}_{\tilde{W}_2^\pm}$) are known.  Then, the momenta of charged
Winos, $\vec{P}_{\tilde{W}_I^\pm}$ with $I=1$ and $2$, are given in
the following form:
\begin{align}
  \vec{P}_{\tilde{W}_I^\pm} = c_I \vec{n}_{\tilde{W}_I^\pm},
\end{align}
with $c_I$ being constants.  The constants $c_1$ and $c_2$ can be
obtained by using the transverse momentum conservation. Concentrating
on events in which the decay products of gluinos contain only hadrons
and Winos, the following relation should be held:
\begin{align}
  \left[
    \sum_{I=1}^2 \vec{P}_{\tilde{W}_I^\pm} + 
    \sum_{j: {\rm jets}} \vec{p}_{j} 
  \right]_T
  = 0,
  \label{sumpt=0}
\end{align}
where the subscript $T$ denotes transverse components.  In our
analysis, we use Eq.\,\eqref{sumpt=0} to calculate $c_1$ and $c_2$, with
which Wino momenta are determined.  With $\vec{P}_{\tilde{W}_I^\pm}$
being given, the reconstructed Wino mass (in association with each
track) is given by
\begin{align}
  m_{\tilde{W}}^{\rm (rec)} = \frac{\sqrt{1-\beta_I^2}}{\beta_I}
  |\vec{P}_{\tilde{W}_I}|,\label{mWino(rec)}
\end{align}
where $\beta_I$ is the measured velocity of the $I$-th charged Wino.

In order to see how well the Wino mass can be reconstructed, we study
the distribution of $m_{\tilde{W}}^{\rm (rec)}$.  For the Wino mass
determination utilizing the above procedure, charged Wino samples with
smaller uncertainties in the boost factor are necessary; this motivates us
to use charged Winos with relatively low velocity.  In addition,
because the ``missing'' momentum evaluated using the jet momenta is
assumed to be compensated by the Wino momenta, events with neutrino
emission should be excluded in this study. Thus, we do not use events
with an isolated lepton\footnote
{A term of lepton used in the selection means electrons and muons.}
 because it may originate from the decay of
$W$-boson.  Thus, for the Wino mass determination, we use the events
satisfying the following conditions (as well as Requirements 1, 2 and
3 in the previous section):
\begin{itemize}
\item[(1a)] There exists a charged Wino with $L_T>10\ {\rm cm}$ and
  $\beta<0.8$.
\item[(1b)] There exists no isolated lepton.\footnote
	    {For the detection and isolation of leptons, we adopt the
	    default condition of \texttt{Delphes} defined in the card
	    \texttt{FCChh.tcl}: a charged lepton is detected with some
	    non-zero tracking efficiency if its momentum satisfies
	    $p_T > 0.5\,\mathrm{GeV}$ and $|\eta| < 6$.  It is also
	    considered to be isolated if the scalar sum of the $p_T$
	    values of particles that surround it within a cone of
	    radius $\sqrt{ \Delta \eta^2 + \Delta \phi^2} < 0.3$ and
	    possess $p_T > 0.5\,\mathrm{GeV}$, divided by its $p_T$,
	    is less than $0.1$ ($0.2$) for an electron (a muon).}
\end{itemize}
We calculate the reconstructed Wino mass $m_{\tilde{W}}^{\rm (rec)}$
for all the charged Winos satisfying (1a).

\begin{figure}
  \centerline{\epsfxsize=0.95\textwidth\epsfbox{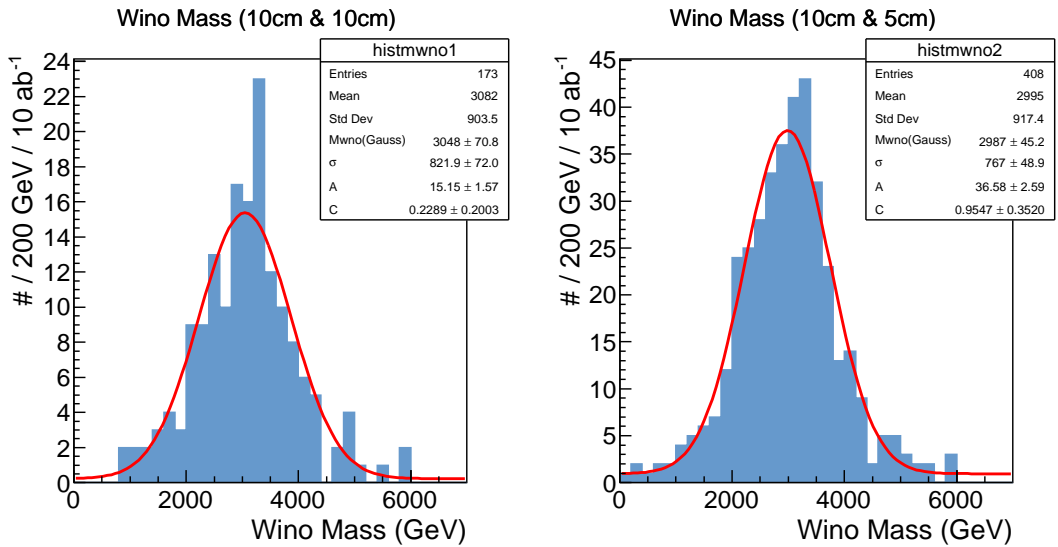}}
  \caption{\small Distribution of the reconstructed Wino mass for
    Sample Point 1, taking $L_{T,2}^{\rm (min)}=10\ {\rm cm}$ (left)
    and $5\ {\rm cm}$ (right) in a single experiment with the
    integrated luminosity of ${\cal L}=10\ {\rm ab}^{-1}$.}
  \label{fig:mwno_mg6tev}
\end{figure}

\begin{figure}
  \centerline{\epsfxsize=0.95\textwidth\epsfbox{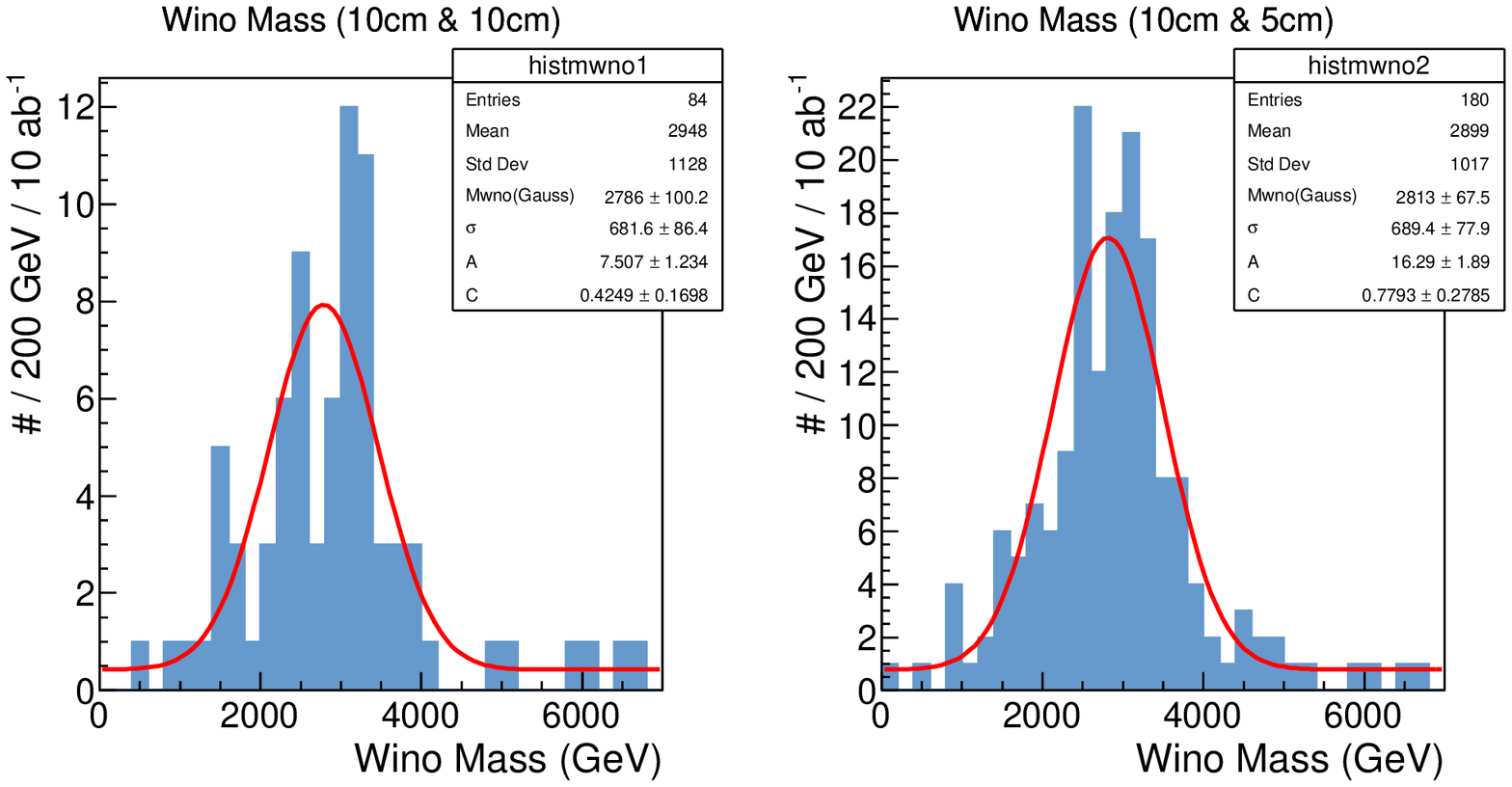}}
  \caption{\small Distribution of the reconstructed Wino mass for
    Sample Point 2, taking $L_{T,2}^{\rm (min)}=10\ {\rm cm}$ (left)
    and $5\ {\rm cm}$ (right) in a single experiment with the
    integrated luminosity of ${\cal L}=10\ {\rm ab}^{-1}$.}
  \label{fig:mwno_mg7tev}
\end{figure}

\begin{figure}
  \centerline{\epsfxsize=0.95\textwidth\epsfbox{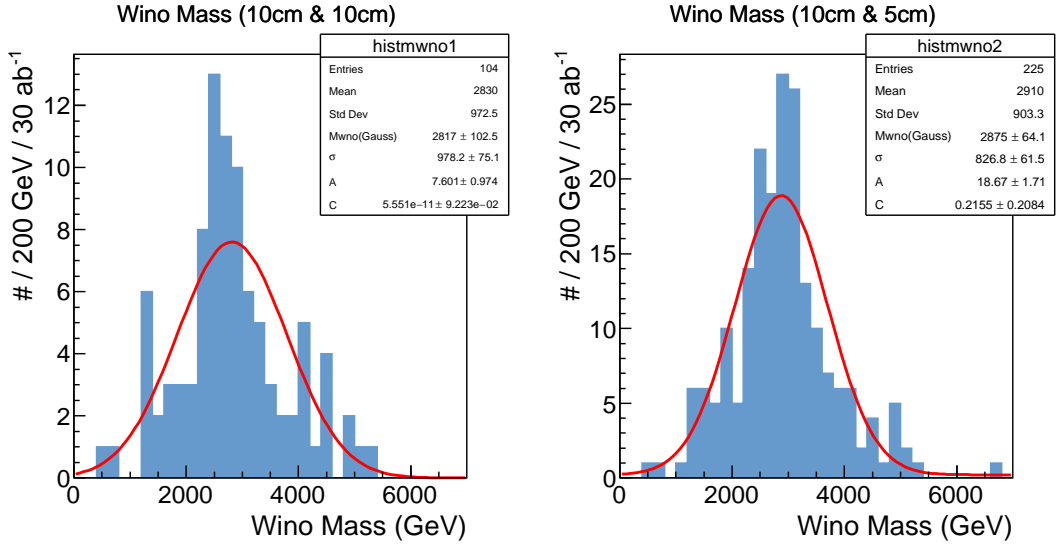}}
  \caption{\small Distribution of the reconstructed Wino mass for
    Sample Point 3, taking $L_{T,2}^{\rm (min)}=10\ {\rm cm}$ (left)
    and $5\ {\rm cm}$ (right) in a single experiment with the
    integrated luminosity of ${\cal L}=30\ {\rm ab}^{-1}$.}
  \label{fig:mwno_mg8tev}
\end{figure}

In figures \ref{fig:mwno_mg6tev}, \ref{fig:mwno_mg7tev}, and
\ref{fig:mwno_mg8tev}, we show the distribution of the reconstructed
Wino mass for Sample Points 1, 2, and 3, respectively, where not only
$L_{T,2}^{\rm (min)}=10\ {\rm cm}$ but also $5\ {\rm cm}$
cases are shown as explained before.
As mentioned earlier, the Wino with $L_{T,2}^{\rm (min)}=5\ {\rm cm}$ 
is not used for the mass determination; it is used to show impact on the mass 
determination from altering the event selections. 
This is also the case for Bino and gluino.
For Sample Points 1 and 2 (Sample Point 3), we take the integrated luminosity
${\cal L}$ of $10\ {\rm ab}^{-1}$ ($30\ {\rm ab}^{-1}$).  The peak of
the distribution is close to the true Wino mass.  In order to
determine the position of the peak, we use a fitting function in the
form of a Gaussian function plus a constant:
\begin{align}
  \frac{d N}{d m^{\rm (rec)}} = 
  A \exp \left[
    - \frac{(m^{\rm (rec)} - \bar{m})^2}{2\sigma^2}
  \right]
  + C,
  \label{fittingfn}
\end{align}
where $\bar{m}$, $\sigma$, $A$, and $C$ are fitting parameters.  In
each figure, the fitting function with the best-fit parameters is
shown in the red line.  The best-fit value of $\bar{m}$, denoted as
{\tt Mwno(Gauss)}, is also shown in each figure.

We expect to extract information about the Wino mass from the position
of the Gaussian peak (i.e.~the peak of the fitting function).  The
position of the peak may not exactly agree with the true Wino mass
even with large enough statistics.  We expect that this is due to our
choice of the fitting function given in eq.~\eqref{fittingfn}, and
that we may find a better fitting function without such a systematic
effect.  The detailed study of the fitting function is beyond the
scope of this paper, and we simply assume that the relation between
characteristic features of the fitting function (like the position of
the peak) and the true mass can be well understood by, for example, a
reliable MC analysis.  In addition, as indicated in the figures, the
statistical uncertainty in the measurement of the Wino mass is accounted for.
Here, we evaluate the statistical uncertainty by using 100 independent
data sets for each Sample Point; we determine the position of the
Gaussian peak for each data set, and the standard deviation of the
peak position is regarded as the statistical uncertainty in the Wino
mass determination.  The statistical uncertainty as well as the mean
of the peak position based on 100 data sets are summarized in table
\ref{table:winomass}.

\begin{table}[t]
  \begin{center}
    \begin{tabular}{|l|ccc|ccc|}
      \hline
      & \multicolumn{3}{c|}{$L_{T,2}^{\rm (min)}=10\ {\rm cm}$}
      & \multicolumn{3}{c|}{$L_{T,2}^{\rm (min)}=5\ {\rm cm}$}
      \\
      & Point 1 & Point 2 & Point 3
      & Point 1 & Point 2 & Point 3
      \\ \hline
      $\langle m_{\tilde{W}} \rangle_{100}$ &
      2922 & 2915 & 2896 &
      2907 & 2891 & 2891
      \\
      $\delta m_{\tilde{W}}$ &
      75 & 109 & 92 &
      44 & 71  & 57
      \\
      \hline
    \end{tabular}
    \caption{The mean of the peak position based on 100 data sets of
   $pp\rightarrow \tilde{g}\tilde{g}$ events for each Sample Point,
   denoted as $\langle m_{\tilde{W}} \rangle_{100}$, as well as the
   uncertainty in the Wino mass determination (in units of GeV).}
   \label{table:winomass}
  \end{center}
\end{table}

So far, we have only considered the $pp\rightarrow\tilde{g}\tilde{g}$
process to determine the Wino mass.  However, if gluino is so heavy that
the cross section for the gluino pair production is significantly
suppressed, it may be more efficient to use the electroweak production
of charged Winos.  In particular, the process
$pp\rightarrow\tilde{W}^{+}\tilde{W}^{-}+\mathrm{jets}$ may be used for
the Wino mass determination, where a large missing transverse momentum
is necessary to determine the Wino momenta as well as to pass the trigger selection.
Here, we study such a possibility assuming that $m_{\tilde{W}}=2.9\,\mathrm{TeV}$ and that
the gluino is out of kinematical reach of the FCC.  With this choice of
the Wino mass, we performed the MC simulation of the processes up to two
hard jets and obtained
$\sigma(pp\rightarrow\tilde{W}^{+}\tilde{W}^{-}+\mathrm{jets})\simeq
0.15\,\mathrm{fb}$ for the missing transverse momentum larger than
$1\,\mathrm{TeV}$.  As for the event selection, we require the same
requirements as the previous analysis, i.e.~Requirements 1, 2, and 3 in
the previous section and conditions (1a) and (1b) in this section.  We
assume that standard-model backgrounds do not exist after requiring two
(long enough) charged Wino tracks.  Then, we determine the reconstructed
Wino mass using eq.~\eqref{mWino(rec)} for each Wino track.

\begin{figure}
 \centerline{\epsfxsize=0.95\textwidth\epsfbox{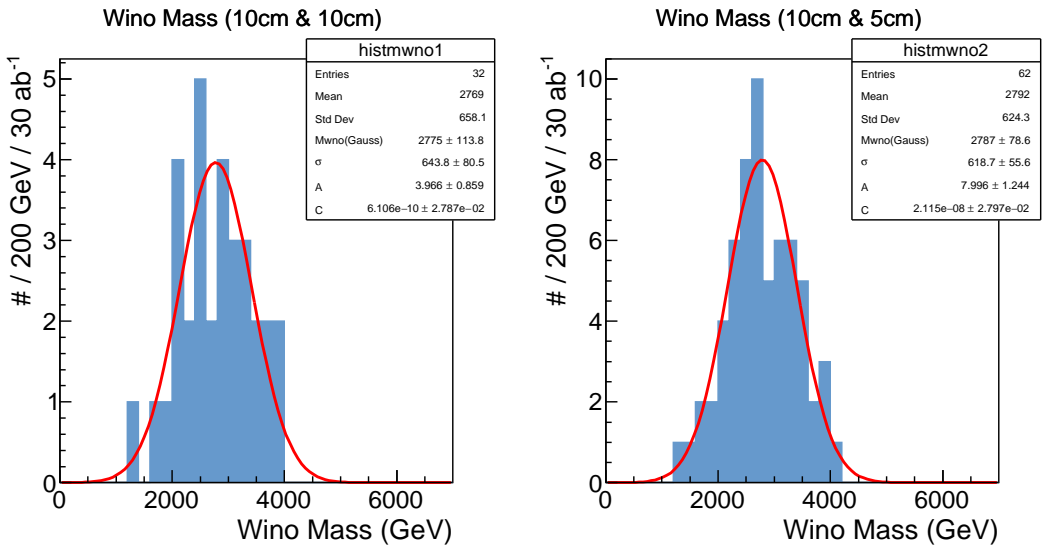}}
 \caption{\small Distribution of the reconstructed Wino mass for the
   chargino pair production process, taking $L_{T,2}^{\rm (min)}=10\
   {\rm cm}$ (left) and $5\ {\rm cm}$ (right) in a single experiment
   with the integrated luminosity of ${\cal L}=30\ {\rm ab}^{-1}$.}
 \label{fig:mwno_Drell_Yan_0}
\end{figure}

\begin{table}[t]
  \begin{center}
    \begin{tabular}{|l|c|c|}
      \hline
      & $L_{T,2}^{\rm (min)}=10\ {\rm cm}$
      & $L_{T,2}^{\rm (min)}=5\ {\rm cm}$
      \\ \hline
      $\langle m_{\tilde{W}} \rangle_{100}$ &
      2751 & 2754
      \\
      $\delta m_{\tilde{W}}$ &
      128 & 84
      \\
      \hline
    \end{tabular}
    \caption{The mean of the peak position based on 100 data sets
   $pp\rightarrow \tilde{W}^{+} \tilde{W}^{-}+\mathrm{jets}$ events,
   denoted as $\langle m_{\tilde{W}} \rangle_{100}$, as well as the
   uncertainty in the Wino mass determination (in units of GeV).}
   \label{table:winomass_Drell_Yan}
  \end{center}
\end{table}

In figure \ref{fig:mwno_Drell_Yan_0}, we show the distribution of the
reconstructed Wino mass, taking the integrated luminosity
$\mathcal{L}$ of $30\,\mathrm{ab}^{-1}$.  We can see that, with
$\mathcal{L}\sim 30\,\mathrm{ab}^{-1}$, we may observe a peak close to
the true Wino mass.  We also perform a fit using the fitting function
given in eq.~\eqref{fittingfn}.  From the figure, one can see the
statistical uncertainty in the fit and the difference between the true Wino mass and
the position of the Gaussian peak.  Although the difference is larger
than in the case of $pp\rightarrow \tilde{g}\tilde{g}$ events, we
again assume that the dependence of the position of the peak on the
true mass can be well understood by a reliable MC analysis and so on.
Under this assumption, we evaluate the statistical uncertainty with
100 independent event sets as we described before and summarize the
result in table \ref{table:winomass_Drell_Yan}.

\subsection{Bino mass determination}

Next, we consider the Bino mass determination.  Here, we use the decay
mode $\tilde{B}\rightarrow\tilde{W}^\pm W^\mp$, which has a sizeable
branching ratio in the present case.  Assuming that the Wino mass is
known (see the previous subsection), the Wino four-momentum can be
determined by using the velocity information (or by the conservation
of the transverse momentum).  Thus, we may reconstruct the Bino mass
by calculating the invariant mass of the $\tilde{W}^\pm$ plus
$W$-boson system, if the decay products of the $W$-boson are
identified.

\begin{figure}
  \centerline{\epsfxsize=0.8\textwidth\epsfbox{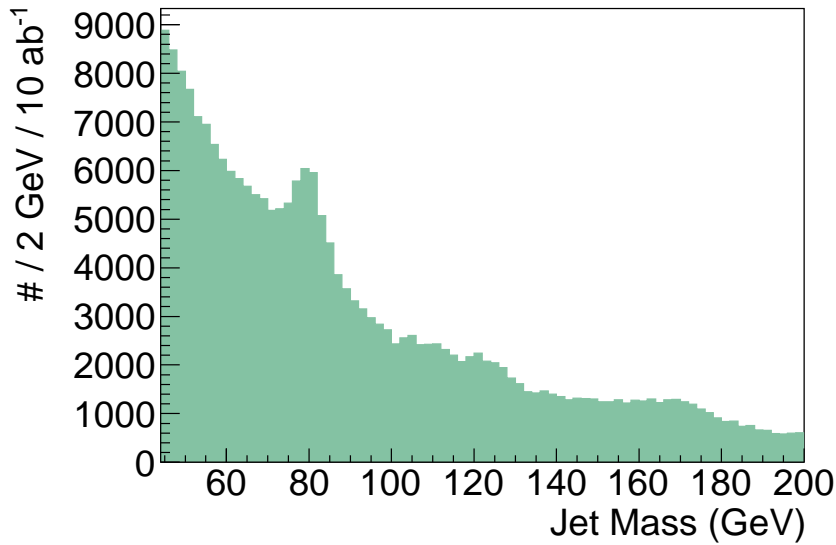}}
  \caption{\small The jet-mass distribution for the Sample Point 1.}
  \label{fig:mjet}
\end{figure}

For this purpose, we use the fact that the $W$-boson produced by the
Bino decay may be highly boosted.  The hadronic decay products of the
boosted $W$-boson may result in a single fat jet with the jet mass
$m_j$ close to the $W$-boson mass.  In figure \ref{fig:mjet}, we show
the distribution of the jet mass for Sample Point 1.  We can observe a
peak at $m_j\sim m_W$.  Thus, if we require $m_j\sim m_W$, we can
obtain $W$-rich samples of jets.  Using such a jet, with denoting its
four-momentum as $p_j$, the reconstructed Bino mass is defined as
\begin{align}
  m_{\tilde{B}}^{\rm (rec)} = 
  \sqrt{(m_{\tilde{W}}^{\rm (rec)} u_{\tilde{W}} + p_j)^2},
  \label{mBino(rec)}
\end{align}
where $u_{\tilde{W}}$ is the four-velocity of the charged Wino:
\begin{align}
  u_{\tilde{W}} = 
  \frac{1}{\sqrt{1-\beta^2}}
  (1, \beta \vec{n}_{\tilde{W}}),
  \label{uvector}
\end{align}
with $\vec{n}_{\tilde{W}}$ being the direction of the Wino momentum
determined by the track information.\footnote
{We may also use the missing momentum information to determine the Wino
  momenta, instead of using the velocity information.  We checked that
  the accuracy of the reconstructed Bino mass is slightly better if
  the Wino momentum is determined by the velocity information.}

Among the events with two charged Wino tracks (i.e.~events that meet
Requirements 1, 2, and 3 in the previous section), those satisfying the
followings are used for the Bino mass reconstruction:\footnote
{One may also use the leptonic decay mode of the $W$-boson, as discussed
  in ref.~\cite{Asai:2008im}.  In such an analysis, the end-point of the
  invariant mass of the Wino plus $\ell^\pm$ system (with $\ell^\pm$
  being a charged lepton) has a sensitivity to the Bino mass.  In the
  present case, however, we found that the analysis with fat jets is
  more powerful for the Bino mass determination.}
\begin{itemize}
\item[(2a)] There exists a charged Wino with $L_T>10\ {\rm cm}$ and
  $\beta<0.9$.  (In order to increase the number of samples, we adopt
  a larger value of the maximal velocity compared to the
  Wino mass determination.)
\item[(2b)] There exists a fat jet with $70\ {\rm GeV} < m_j < 90\
  {\rm GeV}$, with $m_j$ being the jet mass.  We call such a jet as
  ``$W$-jet.''
\end{itemize}
We calculate the reconstructed Bino mass given in
eq.~\eqref{mBino(rec)} for all the pairs of a charged Wino satisfying
(2a) and a $W$-jet satisfying (2b).

\begin{figure}
  \centerline{\epsfxsize=0.95\textwidth\epsfbox{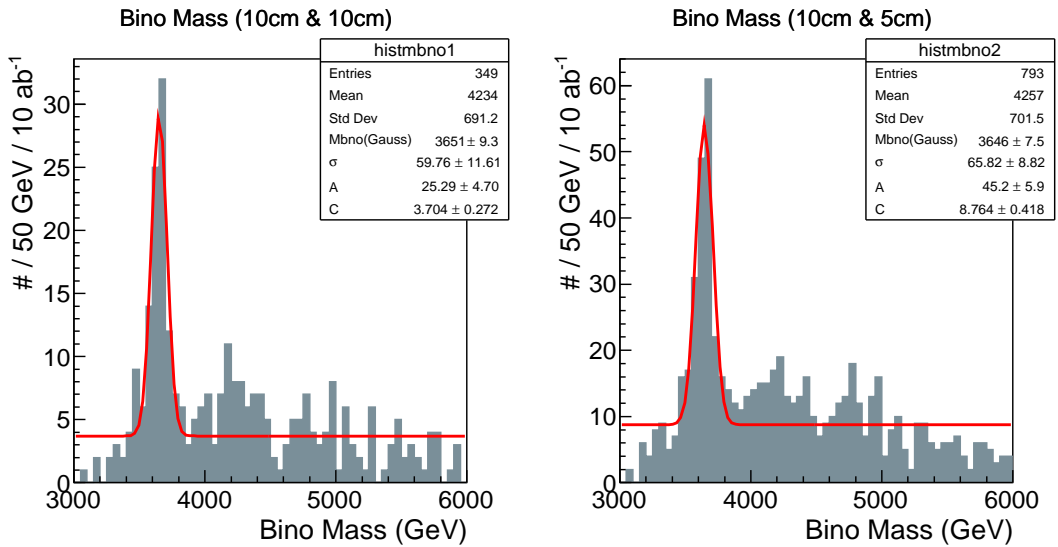}}
  \caption{\small Distribution of $m_{\tilde{B}}^{\rm (rec)}$ for
    Sample Point 1, taking $L_{T,2}^{\rm (min)}=10\ {\rm cm}$ (left)
    and $5\ {\rm cm}$ (right) in a single experiment with the
    integrated luminosity of ${\cal L}=10\ {\rm ab}^{-1}$.}
  \label{fig:mbno_mg6tev}
\end{figure}

\begin{figure}
  \centerline{\epsfxsize=0.95\textwidth\epsfbox{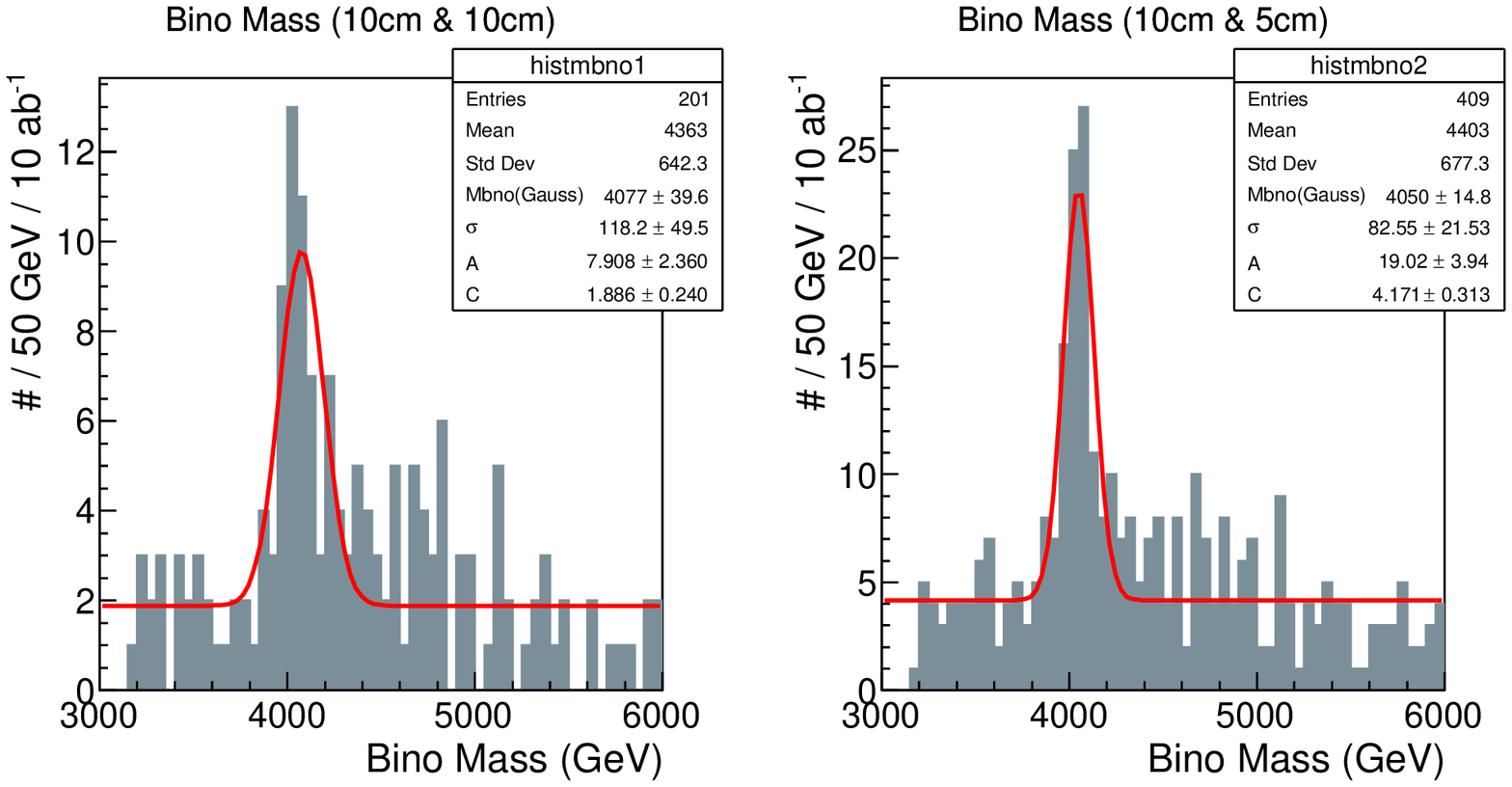}}
  \caption{\small Distribution of $m_{\tilde{B}}^{\rm (rec)}$ for
    Sample Point 2, taking $L_{T,2}^{\rm (min)}=10\ {\rm cm}$ (left)
    and $5\ {\rm cm}$ (right) in a single experiment with the
    integrated luminosity of ${\cal L}=10\ {\rm ab}^{-1}$.}
  \label{fig:mbno_mg7tev}
\end{figure}

\begin{figure}
  \centerline{\epsfxsize=0.95\textwidth\epsfbox{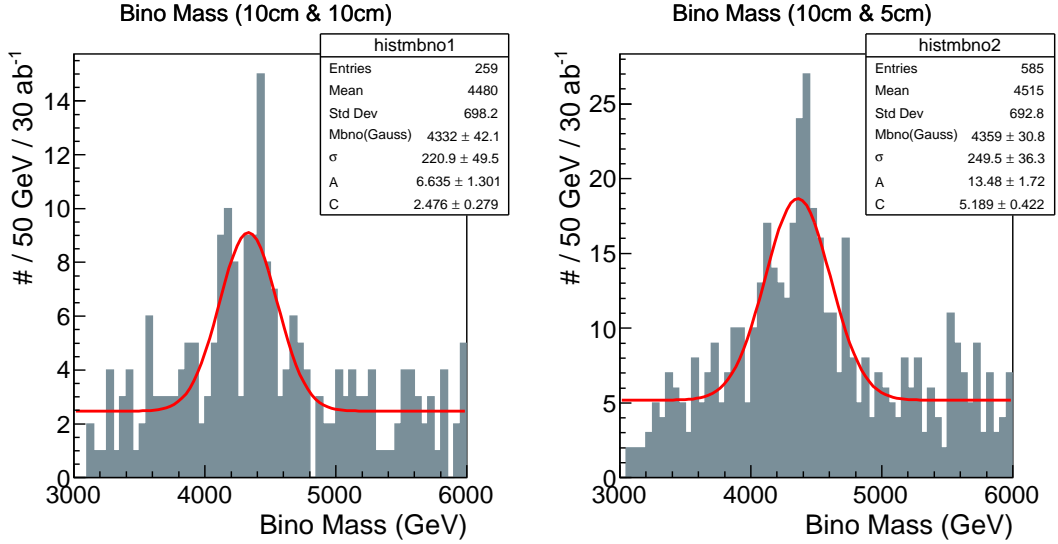}}
  \caption{\small Distribution of $m_{\tilde{B}}^{\rm (rec)}$ for
    Sample Point 3, taking $L_{T,2}^{\rm (min)}=10\ {\rm cm}$ (left)
    and $5\ {\rm cm}$ (right) in a single experiment with the
    integrated luminosity of ${\cal L}=30\ {\rm ab}^{-1}$.}
  \label{fig:mbno_mg8tev}
\end{figure}

In figures \ref{fig:mbno_mg6tev}$\hyphen$\ref{fig:mbno_mg8tev}, we show
the distribution of the reconstructed Bino mass for Sample Points
$1\hyphen 3$, respectively.  Here, the true value of the Wino mass is
used for the calculation of the four-momenta of charged Winos.  We can
see that, in each figure, the peak of the histogram is close to the
true Bino mass and hence the Bino mass determination is possible via
the method using the charged Wino track and the $W$-jet.  We use the
fitting function given in eq.~\eqref{fittingfn} to estimate the
position of the peak.  The fitting function with the best-fit
parameters is shown in the red line.  For each Sample Point, the
best-fit value of the position of the Gaussian peak (denoted as {\tt
Mbno(Gauss)}) and its statistical uncertainty are shown in the figure.

There exist several sources of uncertainties in the reconstructed Bino
mass.  We observe a slight deviation of the position of the Gaussian
peak from the true Bino mass.  We found that the position of the peak
of the fitting function is likely to be lower than the true Bino mass
with the present choice of cut parameters. Such a systematic effect in
the fitting is assumed to be corrected by, for example, a reliable MC
analysis, and hence is not included in the estimation of the
uncertainty in the Bino mass measurement.  The position of the
Gaussian peak should also have a statistical uncertainty.  As in the
case of the Wino mass determination, the statistical uncertainty
(denoted as $\delta m_{\tilde{B}}^{\rm (stat)}$) is determined by
using 100 independent data sets for each Sample Point.  Furthermore,
so far, we have used the true Wino mass as an input for the analysis.
As we discussed in the previous subsection, the Wino mass is also a
parameter that should be experimentally determined, and the
uncertainty in the measured Wino mass becomes a source of the
uncertainty in the reconstructed Bino mass.  In order to see how large
such an effect is, we estimate the change of the reconstructed Bino
mass with respect to the change of the input Wino mass.  With varying
the input Wino mass by $100\ {\rm GeV}$, we found that the
reconstructed Bino mass changes by $\sim 100\ {\rm GeV}$.  Thus, we
take $\delta m_{\tilde{B}}^{(m_{\tilde{W}})}\sim \delta
m_{\tilde{W}}$, where $\delta m_{\tilde{B}}^{(m_{\tilde{W}})}$ is the
uncertainty associated with Wino mass uncertainty $\delta m_{\tilde{W}}$
in the reconstructed Bino mass.  We add all the
uncertainties in quadrature to estimate the expected accuracy of the
Bino mass measurement.  The uncertainties in the Bino mass
reconstruction, as well as the mean of the peak position based on 100
data sets, are summarized in table \ref{table:binomass}.

\begin{table}[t]
  \begin{center}
    \begin{tabular}{|l|ccc|ccc|}
      \hline
      & \multicolumn{3}{c|}{$L_{T,2}^{\rm (min)}=10\ {\rm cm}$}
      & \multicolumn{3}{c|}{$L_{T,2}^{\rm (min)}=5\ {\rm cm}$}
      \\
      & Point 1 & Point 2 & Point 3
      & Point 1 & Point 2 & Point 3
      \\ \hline
      $\langle m_{\tilde{B}} \rangle_{100}$ &
      3651 & 4036 & 4394 &
      3651 & 4036 & 4397
      \\
      $\delta m_{\tilde{B}}^{\rm (stat)}$ &
      15 & 36 & 42 &
      10 & 25 & 29
      \\
      $\delta m_{\tilde{B}}^{(m_{\tilde{W}})}$ &
      75 & 109 & 92 &
      44 & 71  & 57
      \\
      $\delta m_{\tilde{B}}$ &
      76 & 115 & 101 &
      45 & 75  & 64
      \\
      \hline
    \end{tabular}
    \caption{The mean of the peak position based on 100 data sets for
      each Sample Point, denoted as $\langle m_{\tilde{B}}
      \rangle_{100}$, as well as the uncertainty in the Bino mass
      determination (in units of GeV).  The total uncertainty, $\delta
      m_{\tilde{B}}$, is obtained by adding $\delta m_{\tilde{B}}^{\rm
        (stat)}$ and $\delta m_{\tilde{B}}^{(m_{\tilde{W}})}$ in
      quadrature.}
    \label{table:binomass}
  \end{center}
\end{table}

\subsection{Gluino mass reconstruction}

The last subject is the gluino mass reconstruction.  We consider the
gluino pair production and use the hemisphere analysis to separate
the decay products of one gluino from those of the other.  The
invariant mass of each hemisphere is regarded as the reconstructed
gluino mass.  We need to determine the Wino momenta by using the
conservation of the transverse momentum (see eq.~\eqref{sumpt=0}).
Thus, we should use events in which all the decay products of gluinos
are detected.  In order to eliminate events which might contain
neutrino emission, we do not use events containing isolated leptons.
We use the events that satisfy the next condition:
\begin{itemize}
\item[(3a)] There is no isolated lepton.
\end{itemize}

\begin{figure}
  \centerline{\epsfxsize=0.95\textwidth\epsfbox{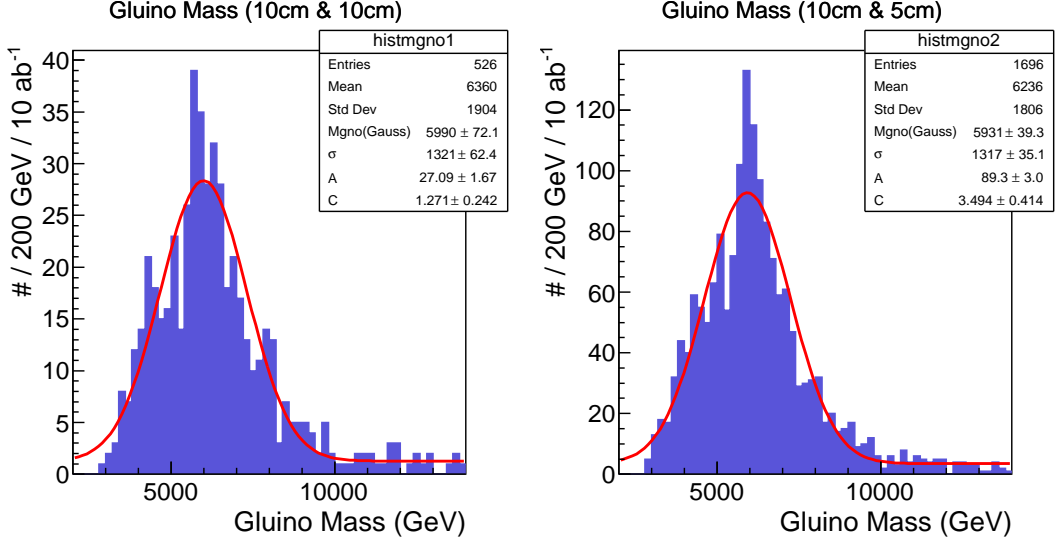}}
  \caption{\small Distribution of $m_{\tilde{g}}^{\rm (rec)}$ for
    Sample Point 1, taking $L_{T,2}^{\rm (min)}=10\ {\rm cm}$ (left)
    and $5\ {\rm cm}$ (right) in a single experiment with the
    integrated luminosity of ${\cal L}=10\ {\rm ab}^{-1}$.}
  \label{fig:mgno_mg6tev}
\end{figure}

\begin{figure}
  \centerline{\epsfxsize=0.95\textwidth\epsfbox{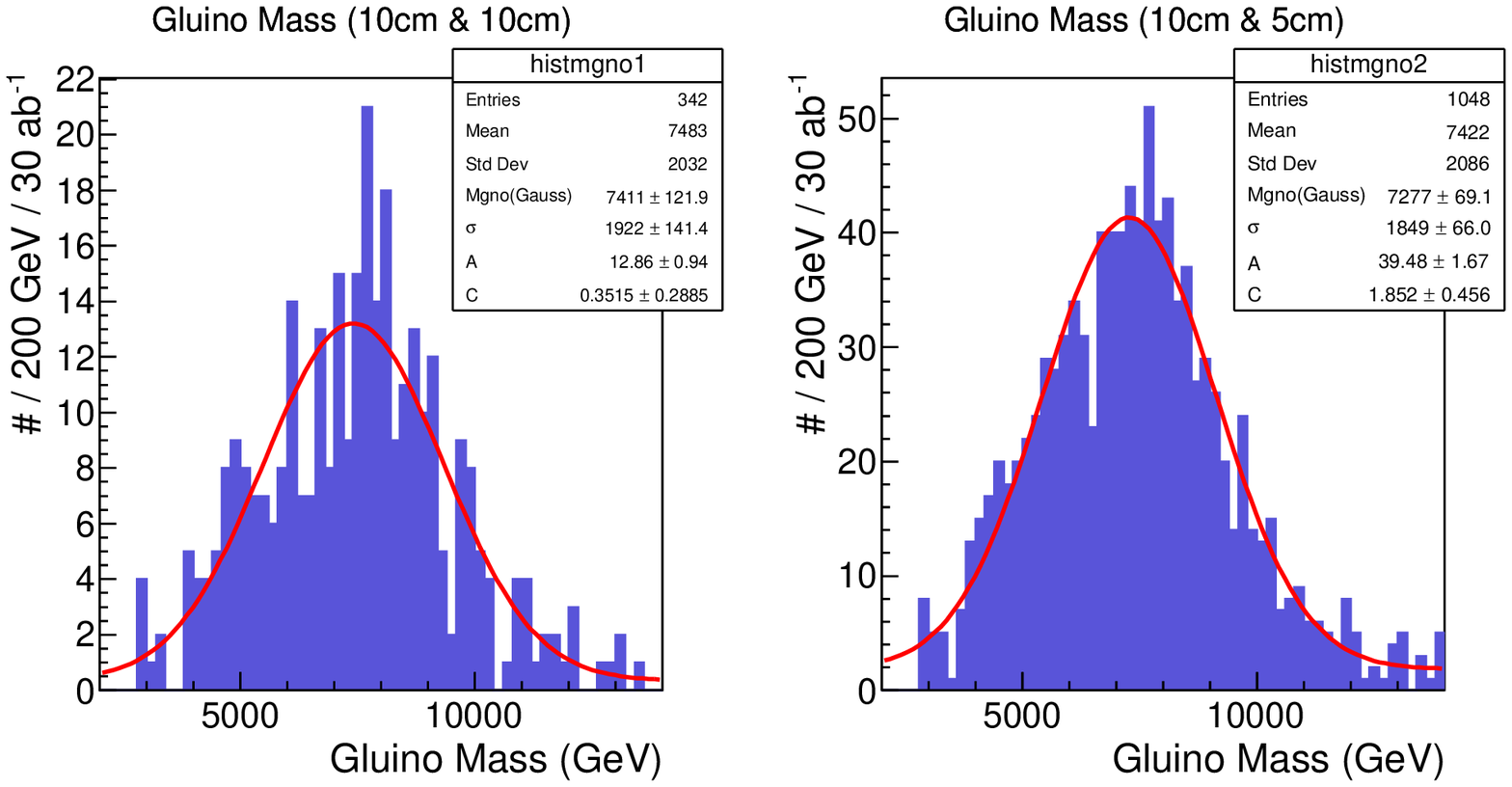}}
  \caption{\small Distribution of $m_{\tilde{g}}^{\rm (rec)}$ for
    Sample Point 3, taking $L_{T,2}^{\rm (min)}=10\ {\rm cm}$ (left)
    and $5\ {\rm cm}$ (right) in a single experiment with the
    integrated luminosity of ${\cal L}=30\ {\rm ab}^{-1}$.}
  \label{fig:mgno_mg8tev}
\end{figure}

In the hemisphere analysis for the gluino mass reconstruction, two
charged Winos $\tilde{W}^\pm_{1,2}$ and high-$p_T$ jets are assigned
to one of two hemispheres, $H_1$ or $H_2$.  The assignment is
performed as follows:
\begin{enumerate}
\item Two charged Winos are assigned to different hemispheres:
  \begin{align}
    \tilde{W}^\pm_I \in H_I~~~(I=1,2).
  \end{align}
\item For each high-$p_T$ jet $j$ with $|\eta|<2$ (with $\eta$ being
  pseudo-rapidity of the jet),
  \begin{align}
    \left\{ \begin{array}{ll}
        j \in H_1 & :~ \mbox{if}\ 
        d (p_{H_1}, p_j) < d (p_{H_2}, p_j)\\
        j \in H_2 & :~ \mbox{if}\
        d (p_{H_2}, p_j) < d (p_{H_1}, p_j)
      \end{array} \right. ,
  \end{align}
  where $p_j=(E_j,\vec{p}_j)$ is the four-momentum of $j$, and
  $p_{H_I}=(E_{H_I},\vec{p}_{H_I})$ is the four-momentum of the $I$-th
  hemisphere that is defined as
  \begin{align}
    p_{H_I} = P_{\tilde{W}^{\pm}_I} + \sum_{j \in H_I} p_j.
    \label{P_HI}
  \end{align}
  The function $d (p_{H}, p_j)$ is given by
  \begin{align}
    d (p_{H}, p_j) =
    \frac{(E_{H} - |\vec{p}_{H}| \cos\theta_{Hj}) E_{H}}
    {(E_{H} + E_j)^2},
  \end{align}
  with $\theta_{Hj}$ being the angle between $\vec{p}_{H}$ and
  $\vec{p}_j$ \cite{Ball:2007zza}.  

\item[3.] Jets with $|\eta|>2$ are not used for the analysis.
\end{enumerate}
In our MC analysis, we determine the hemispheres by iteration.  In the
first step of the iteration, we take $p_{H_I} =
P_{\tilde{W}^{\pm}_I}$, and all the high-$p_T$ jets are merged to one
of the hemispheres which gives smaller $d (p_{H_I}, p_j)$.  Then,
after the first step, we use Eq.\ \eqref{P_HI} and repeat the
iteration with re-calculating $p_{H_I}$.  Compared to models in which
particles responsible for the missing momentum are totally invisible,
the hemisphere analysis in the present model is easier.  This is
because the momentum information of all the final state particles is
available after reconstructing the Wino momenta.  Once the hemispheres
are determined, the reconstructed gluino mass for each hemisphere is
defined as
\begin{align}
  m_{\tilde{g}}^{\rm (rec)} = 
  \sqrt{p_{H_I}^2}.
  \label{mgluino(rec)}
\end{align}
In figures \ref{fig:mgno_mg6tev}$\hyphen$\ref{fig:mgno_mg8tev}, we show the
distribution of the reconstructed gluino mass; in the figures, the
true Wino mass is used to determine the four-momenta of Winos.

\begin{figure}
  \centerline{\epsfxsize=0.95\textwidth\epsfbox{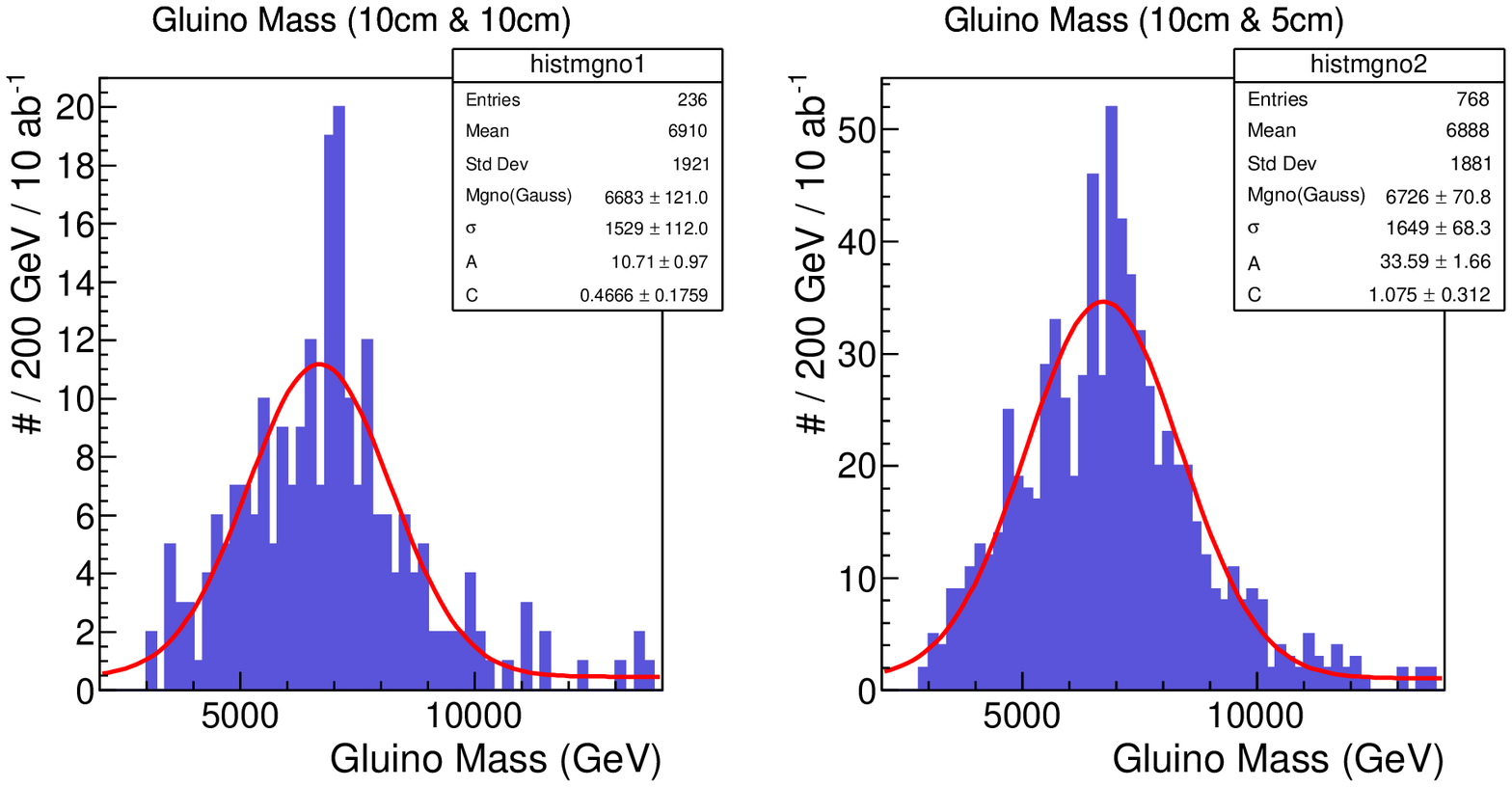}}
  \caption{\small Distribution of $m_{\tilde{g}}^{\rm (rec)}$ for
    Sample Point 2, taking $L_{T,2}^{\rm (min)}=10\ {\rm cm}$ (left)
    and $5\ {\rm cm}$ (right) in a single experiment with the
    integrated luminosity of ${\cal L}=10\ {\rm ab}^{-1}$.}
  \label{fig:mgno_mg7tev}
\end{figure}

We can see that the position of the Gaussian peak may have a sizeable
deviation from the true gluino mass, although the position of the peak
has a positive correlation with the true gluino mass.  The deviation
is partly a systematic effect due to our choices of the fitting
function and cut parameters. Such a systematic effect is assumed to be
understood well so that it does not affect the accuracy of the
observed gluino mass.  The statistical uncertainty, $\delta
m_{\tilde{g}}^{\rm (stat)}$, is estimated by using 100 independent set
of events for each Sample Point.  Another source of the error in the
gluino mass reconstruction is the Wino mass.  In our procedure of
determining the gluino mass, the Wino mass is needed as an input parameter
to determine the four-momenta of charged Winos.  In order to estimate
the effect of the Wino mass uncertainty, we vary the input Wino mass
by $100\ {\rm GeV}$ and find that the reconstructed gluino mass
changes by $\sim 100\ {\rm GeV}$.  Thus, the uncertainty in the
reconstructed gluino mass due to the Wino mass uncertainty is
estimated to be $\delta m_{\tilde{g}}^{(m_{\tilde{W}})}\sim \delta
m_{\tilde{W}}$.  Adding the uncertainties in quadrature, we estimate the
expected accuracy of the gluino mass determination; in table
\ref{table:gluinomass}, we summarize the uncertainties in the gluino
mass reconstruction, as well as the mean of the peak position based on
the analysis with 100 independent data sets for each Sample Point.

\begin{table}[t]
  \begin{center}
    \begin{tabular}{|l|ccc|ccc|}
      \hline
      & \multicolumn{3}{c|}{$L_{T,2}^{\rm (min)}=10\ {\rm cm}$}
      & \multicolumn{3}{c|}{$L_{T,2}^{\rm (min)}=5\ {\rm cm}$}
      \\
      & Point 1 & Point 2 & Point 3
      & Point 1 & Point 2 & Point 3
      \\ \hline
      $\langle m_{\tilde{g}} \rangle_{100}$ &
      6018 & 6784 & 7473 &
      5988 & 6754 & 7439
      \\
      $\delta m_{\tilde{g}}^{\rm (stat)}$ &
      66 & 101 & 119 &
      33 & 56 & 66
      \\
      $\delta m_{\tilde{g}}^{(m_{\tilde{W}})}$ &
      75 & 109 & 92 &
      44 & 71  & 57
      \\
      $\delta m_{\tilde{g}}$ &
      99 & 149 & 150 &
      55 & 90 & 87
      \\
      \hline
    \end{tabular}
    \caption{The mean of the peak position based on 100 data sets for
      each Sample Point, denoted as $\langle m_{\tilde{g}}
      \rangle_{100}$, as well as the uncertainty in the gluino mass
      determination (in units of GeV).  The total uncertainty, $\delta
      m_{\tilde{g}}$, is obtained by adding $\delta m_{\tilde{g}}^{\rm
        (stat)}$ and $\delta m_{\tilde{g}}^{(m_{\tilde{W}})}$ in
      quadrature.}
    \label{table:gluinomass}
  \end{center}
\end{table}

Before closing this subsection, we comment on another possibility to
determine the gluino mass, i.e., the gluino mass determination from
the cross section measurement, although a detailed study of such a
possibility is beyond the scope of this paper.  The cross section of
the gluino pair production is highly dependent on the gluino mass.
Consequently, if the cross section of the gluino pair production
process can be measured, and also if the cross section can be
theoretically well-understood, gluino mass can be determined.  The
dominant sources of uncertainties are the parton distribution function
(PDF) uncertainty and the uncertainty in the MC simulation (see, for example,
ref.~\cite{Aaboud:2017vwy} for typical sources of MC simulation
uncertainty).  We expect that both of them give $\sim 10\ \%$
uncertainties in the theoretical calculation of the cross section,
based on the current estimate of the PDF uncertainty in
ref.~\cite{Butterworth:2015oua}.  Because the cross section scales as
$\sim m_{\tilde{g}}^{-(7\text{--}8)}$ for the range of the gluino mass of our
interest (see Table \ref{table:samplept}), the uncertainty in the
gluino mass determination using the cross section is estimated to be a
few \%.  We also note that, from the consistency between the gluino
mass from the hemisphere analysis and that from the cross section
information, we can check if the particle produced in the pair
production event is consistent with gluino (i.e., a Majorana fermion
in the adjoint representation of $SU(3)_C$), which provides an
important test of SUSY model.

\section{Implications}
\label{sec:implications}
\setcounter{equation}{0}

Let us discuss the implications of the gaugino mass determination.  In
the PGM model, the gaugino masses depend on the following three
fundamental parameters:
\begin{align}
  m_{3/2},~~~|L|,~~~\phi_L \equiv \mbox{arg} (L).
\end{align}
(Remember that $m_{3/2}$ is real and positive in our convention.)
Even though there are three fundamental parameters for three gaugino
masses, there is a non-trivial constraint on the gaugino masses in PGM
model; based on eqs.~\eqref{m1}$\hyphen$\eqref{m3}, we can find
\cite{Asai:2007sw}
\begin{align}
  \left| \frac{10 g_1^{2}}{3g_3^{2}} |M_3 (M_\mathrm{S})|
    - \frac{g_1^{2}}{g_2^{2}} |M_2 (M_\mathrm{S})| \right|
  \lesssim
  |M_1 (M_\mathrm{S})|
  \lesssim
  \frac{10 g_1^{2}}{3g_3^{2}} |M_3 (M_\mathrm{S})|
  + \frac{g_1^{2}}{g_2^{2}} |M_2 (M_\mathrm{S})|.
  \label{PGMrelation}
\end{align}
Thus, by checking if the gaugino masses obey the above relation, it
provides a test of the PGM model.  Simultaneously, with the gaugino
mass measurements, $m_{3/2}$, $|L|$, and $\phi_L$ can be determined.

\begin{figure}[t]
 \begin{center}
  \includegraphics[width=0.6\hsize]{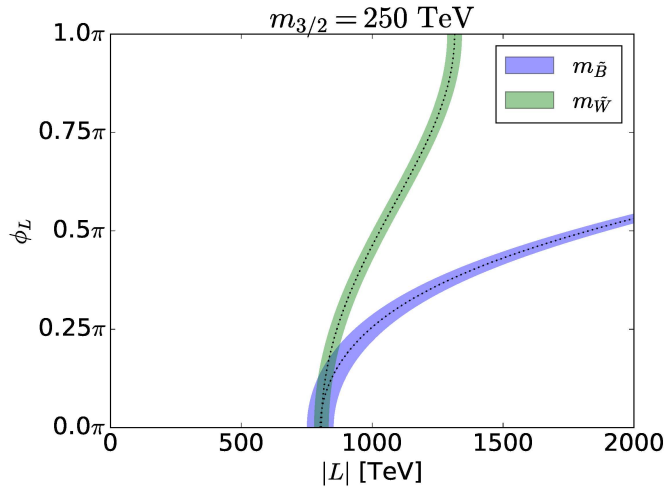}
 \end{center}
 \caption{Contours of constant Bino and Wino masses on the $|L|$ vs.\
   $\phi_L$ plane, taking $m_{3/2} = 250\,\mathrm{TeV}$.  The dotted
   lines show the true Bino and the Wino masses for Sample Point 1,
   while the bands show expected uncertainties due to the uncertainty in
   the gaugino mass determination.}  \label{fig:cont_L_vs_phi}
\end{figure}

\begin{figure}
  \begin{center}
    \includegraphics[width=0.6\hsize]{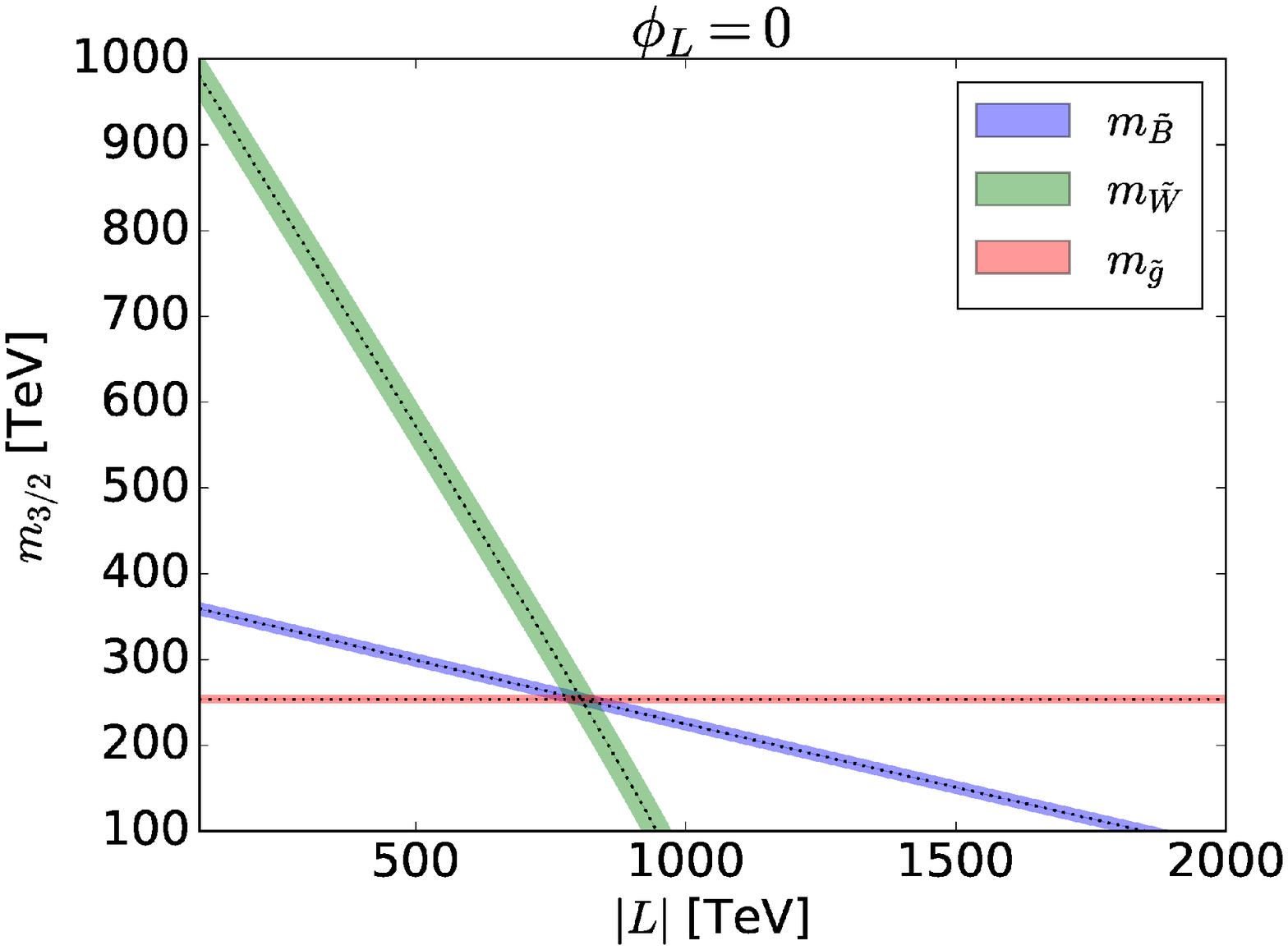}
    \caption{Contours of constant gaugino masses on the $|L|$ vs.\
    $m_{3/2}$ plane, taking $\phi_L=0$.  The dotted lines show the true
    gaugino masses for Sample Point 1, while the bands show expected
    uncertainties due to the uncertainty in the gaugino mass determination.}
    \label{fig:cont_L_vs_m32}
    \vspace{5mm}
    \includegraphics[width=0.6\hsize]{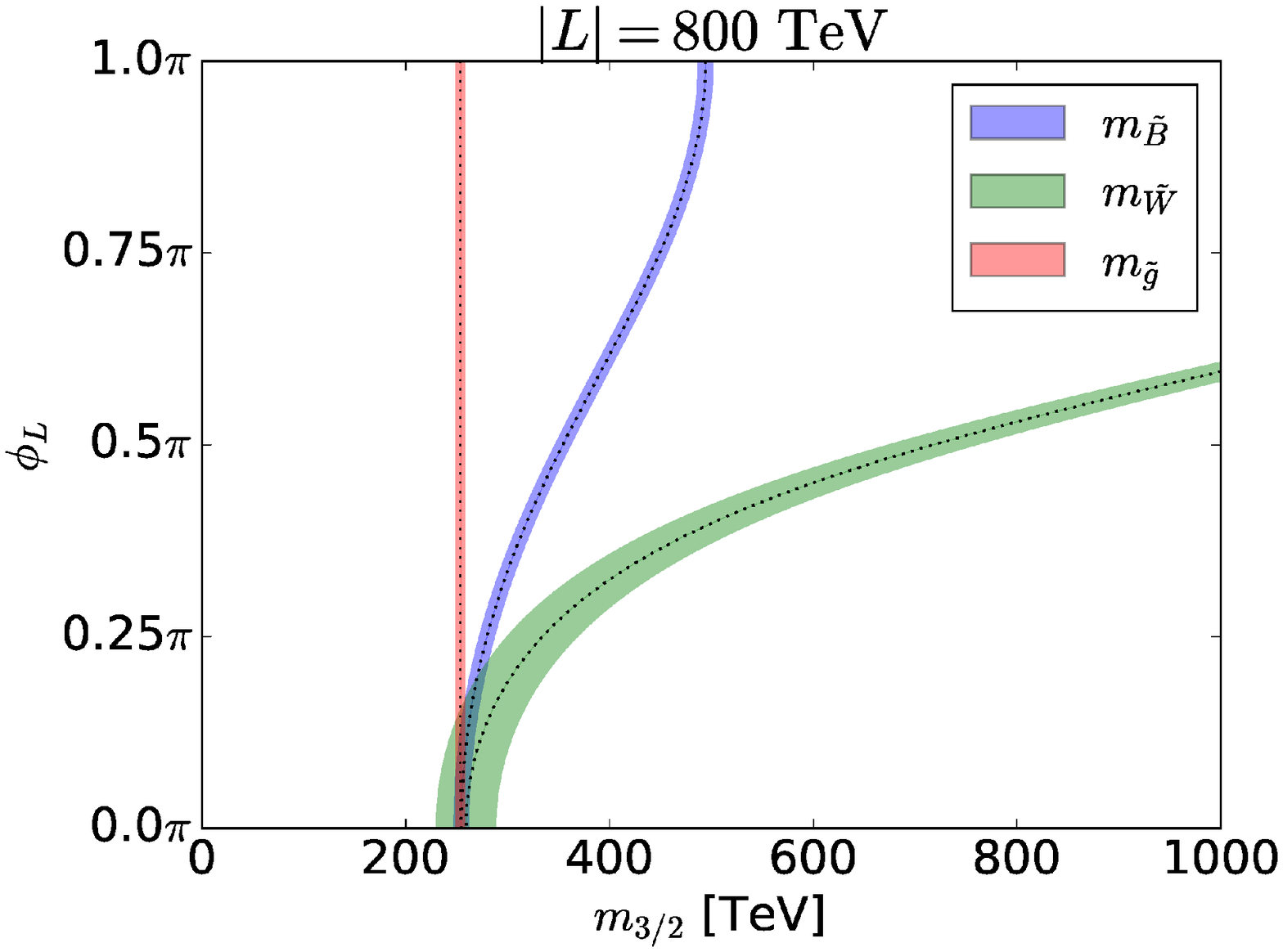}
    \caption{Contours of constant gaugino masses on the $m_{3/2}$ vs.\
    $\phi_L$ plane, taking $|L|=800\,\mathrm{TeV}$.  The dotted lines
    show the true gaugino masses for Sample Point 1, while the bands
    show expected uncertainties due to the uncertainty in the gaugino mass
    determination.}  \label{fig:cont_m32_vs_phi}
  \end{center}
\end{figure}

Here, we consider how well we can test the PGM prediction and how well
we can determine fundamental parameters.  For this purpose, we
consider the parameter region consistent with observed gaugino masses,
adopting Sample Point 1 as an example.  To begin with, we plot
contours of constant gaugino masses on the $|L|$ vs.\ $\phi_L$ plane
in figure \ref{fig:cont_L_vs_phi}, with fixing $m_{3/2}=250\ {\rm GeV}$
(i.e.~the true value of the gravitino mass); the dotted lines are
contours of true Bino and Wino masses, while the bands show the
expected uncertainties given in tables \ref{table:winomass} and
\ref{table:binomass}.  Notice that the gluino mass is independent of
$|L|$ and $\phi_L$, and is $6\ {\rm TeV}$ for $m_{3/2}=250\ {\rm
GeV}$.  We can estimate the uncertainties in the determination of these two
parameters as $\delta|L| \sim 25\,\mathrm{TeV}$ and $\delta \phi_L
\sim \pi/5$.\footnote{
Strictly speaking, there is a correlation among errors of gaugino mass
determinations (see $\delta m_{\tilde{B}}^{(m_{\tilde{W}})}$ in table
\ref{table:binomass} for example).  Here, we just interpret the
overlapped region of the bands as the allowed region and thus obtain a
conservative estimation for uncertainties in model parameters.
}
In figures \ref{fig:cont_L_vs_m32} and \ref{fig:cont_m32_vs_phi}, we
also show the contours of constant gaugino masses on the $|L|$ vs.\
$m_{3/2}$ plane and the $m_{3/2}$ vs.\ $\phi_L$ plane, respectively.  In
each figure, the remaining parameter ($\phi_L$ or $|L|$) is fixed to its
input value; the dotted lines show the contours of true gaugino masses,
while the bands stand for the expected uncertainty due to the uncertainties in
the gaugino mass determination.  Particularly from the gluino mass
information, we can see that the gravitino mass $m_{3/2}$ can be
determined with the accuracy of $\delta m_{3/2} \sim 5\,\mathrm{TeV}$.

Once $|L|$ is determined, it will provide information about the masses
of heavier MSSM particles.  One of the important purposes of high
energy colliders is to understand the mass scales of unknown particles;
determination of $|L|$ gives information about the mass scales of
Higgsinos and heavy Higgses.  From eq.~\eqref{eq:L_def}, we can see
that $|L|$ depends on $|\mu|$, $m_A$, and $\tan\beta$.  In the PGM
model, the masses of the Higgsinos and heavy Higgses are expected to
be of the same order of magnitude, and hence information about $|L|$
gives a good estimate of their mass scale as $|\mu|\sim m_A\sim
|L|/\sin 2\beta$.  More precisely, we may estimate an upper bound on
the mass scale below which Higgisnos or heavy Higgses, whichever
lighter, should exist.  Such an upper bound depends on the hierarchy
between $|\mu|$ and $m_A$ parameterized by
\begin{align}
  x = \frac{m_A}{|\mu|}.
\end{align}
Varying $x$ within the range $x_{\rm min}\leq x\leq x_{\rm max}$, we
calculate the maximal possible value of $\mbox{min} (|\mu|, m_A)$ to
derive the upper bound on the mass scale:
\begin{align}
  \overline{\mbox{min}}(|\mu|, m_A) \equiv 
  \underset{x_{\rm min}\leq x \leq x_{\rm max}}{\mbox{max}}
  \mbox{min} (|\mu|, m_A).
\end{align}
Here, $x_{\rm min}$ and $x_{\rm max}$ parameterize the possible
hierarchy between $|\mu|$ and $m_A$, and both of them are expected to
be of $O(0.1-1)$ in the PGM model.  Based on eq.~\eqref{eq:L_def},
$\overline{\mbox{min}}(|\mu|, m_A)$ is given by
\begin{align}
  \overline{\mbox{min}}(|\mu|, m_A) = 
  \frac{|L|}{\sin 2\beta}
  \frac{1-x_{\rm min}^2}{x_{\rm min}\ln x_{\rm min}^{-2}},
\end{align}
where we consider the case of $x_{\rm min}\leq 1$.

\begin{figure}[t]
 \begin{center}
  \includegraphics[width=0.6\hsize]{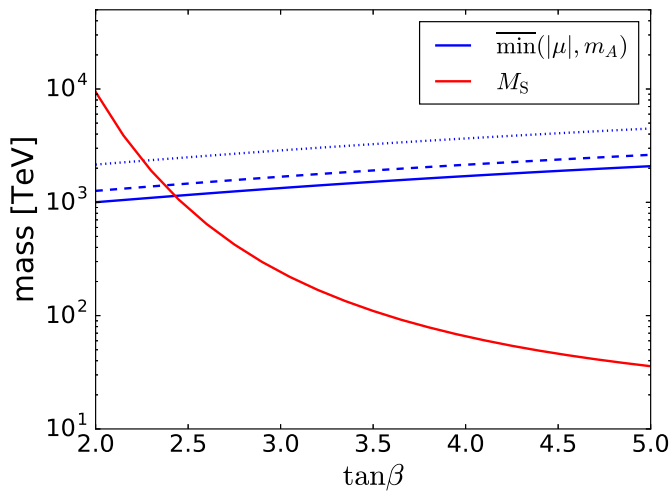}
 \end{center}
 \caption{$\overline{\mbox{min}}(|\mu|, m_A)$ as a function of $\tan\beta$ taking the input value
   of $|L|$ ($=800\,\mathrm{TeV}$) of Sample Point 1 (blue lines).  Solid, dashed, and dotted
   blue lines correspond to $x_{\rm min}=1$, $0.3$, and $0.1$,
   respectively.  $M_\mathrm{S}$ is also shown in the red line.}
 \label{fig:tanb_corr}
\end{figure}

In figure \ref{fig:tanb_corr}, we plot $\overline{\mbox{min}}(|\mu|,
m_A)$ taking the input value of $|L|$ ($=800\,\mathrm{TeV}$) of Sample Point 1.  Here, we
take $x_{\rm min}=0.1$, $0.3$, and $1$.  The upper bound on the mass
scale of Higgsinos and heavy Higgses becomes larger for a larger value
of $\tan\beta$.  We comment here that $\tan\beta$ is also correlated
with the sfermion mass scale $M_\mathrm{S}$ (in particular that of
stops) through the observed value of the Higgs mass.  At the sfermion
mass scale, the Higgs quartic coupling $\lambda$ is determined
as~\cite{Bernal:2007uv, Giudice:2011cg}
\begin{align}
  \lambda(M_\mathrm{S}) = 
  \frac{g_1^2 (M_\mathrm{S}) + g_2^2 (M_\mathrm{S})}{4}
  \cos^2 2\beta +
  \delta \lambda,
  \label{eq_lambda_at_ms}
\end{align}
where $\delta\lambda$ denotes the threshold correction from sfermions
and Higgsinos.  After taking into account the renormalization group
effects, the value of $\lambda$ at the electroweak scale determines
the Higgs mass~\cite{Buttazzo:2013uya}.  In figure \ref{fig:tanb_corr},
we also show $M_\mathrm{S}$ as a function of $\tan\beta$ to realize
the observed Higgs mass of
$m_h=125.18\,\mathrm{GeV}$~\cite{Tanabashi:2018oca}.  As is well
known, $M_\mathrm{S}$ becomes larger as $\tan\beta$ becomes smaller;
this is due to the cancellation between the $\tan\beta$ dependence of
the tree-level contribution to $\lambda$ and the renormalization group
effect on $\lambda$.  Resultantly, $|L|/\sin 2\beta$ and
$M_\mathrm{S}$ have opposite correlation with $\tan\beta$.  Then, two
lines in figure \ref{fig:tanb_corr} intersect with each other at the
mass scale $\sim 10^3\,\mathrm{TeV}$.  This implies that, with the
determination of $|L|$, we will acquire the mass scale below or at
which a new particle (i.e.~Higgsinos, heavy Higgs bosons, or
sfermions) should exist.  Notice that, since the qualitative behaviours
of these two lines always hold, such a mass scale can be always
acquired with the determination of $|L|$.

\section{Conclusions and discussion}
\label{sec:conclusions}
\setcounter{equation}{0}

In this paper, we have discussed a prospect of the study of a SUSY
model at the FCC.  We have concentrated on the so-called PGM model of
SUSY breaking, in which scalars in the MSSM sector have masses of
$O(100)\ {\rm TeV}$ while gauginos are within the kinematical reach of
the FCC, and have studied the SUSY signals at the FCC.  We have paid
particular attention to the case where Wino is lighter than other
gauginos and neutral Wino is the LSP.  In such a case, the charged
Wino has its lifetime of $c\tau_{\tilde{W}^\pm}\simeq 5.75\ {\rm cm}$
and can give unique disappearing tracks in the inner pixel detector.
The charged Wino tracks can be used not only for the reduction of
standard-model backgrounds but also for the reconstruction of SUSY
events.

In the model of our interest, gauginos are the only SUSY particles
that are accessible with the FCC.  We have proposed procedures to
determine their masses.
\begin{itemize}
\item Once two charged Wino tracks are identified, the Wino mass can
  be determined by combining velocity and momentum information of each
  Wino; here, the velocity can be determined by using the time
  information from the pixel detector while the momentum of the Wino
  is determined by using the conservation of the transverse momentum.
\item The Bino mass can be determined by using the fact that the decay
1  process $\tilde{B}\rightarrow\tilde{W}^\pm W^\mp$ has sizeable
  branching ratio.  In particular, at the FCC, the $W$-boson produced
  by the Bino decay can be highly boosted, and hence hadrons from the
  decay of $W$-boson may result in a single fat jet with the jet mass
  close to the $W$-boson mass.  Studying the invariant mass of the
  system consisting of a charged Wino (observed as a disappearing
  track) and a fat $W$-jet, information about the Bino mass can be
  obtained.
\item For the gluino mass determination, we use the fact that,
  requiring two charged Wino tracks in the event, we can observe all
  the decay products of gluinos, i.e.~charged Winos and jets,
  assuming that all the decay products of gluinos are hadrons and
  Winos.  We have discussed the possibility to determine the gluino
  mass using the hemisphere analysis to recombine decay products of
  each gluino.  A distribution of the invariant masses of hemispheres
  gives information about the gluino mass.
\end{itemize}

We emphasize here that there are several assumptions made in this paper.
Charged Wino tracks can be reconstructed with high efficiency, 
even though the transverse flight length is as short as $\sim 10\ {\rm cm}$.
The velocity of charged Wino can be determined with the accuracy 
of $O(1)\ \%$, which depends on the hit-level time resolution of the pixel detector.
For the background, we consider only fake disappearing tracks because 
they are expected to be dominant in high pile-up conditions, 
but it should be kept in mind that physical 
background due to particle scattering also exists to some extent. 
The gaugino mass determination will be affected by contributions 
from additional energies deposited in jets such as pile-ups, but 
such contributions could be sufficiently suppressed by using 
pile-up mitigation techniques, as indicated in ref.~\cite{FCChh_CDR}.
The event selections that we have used in this paper are sufficient to characterize 
the properties of signal events while they might need to be altered if 
more realistic background is considered.
All these conditions may affect the gaugino mass determination, and 
the details in realistic situation will be studied in the future.

The proposed method in this paper is contingent on the existence of
an inner tracking detector with the hit-time resolution of
$O(10)\ {\rm ps}$. For understanding this SUSY model with the
proposed method, technology of the timing-capable sensors
in a high radiation environment has be developed.

Once three gaugino masses are known, the result can be used to
understand the underlying theory of SUSY breaking and to reconstruct
fundamental parameters behind the gaugino masses.  In the PGM model of
our interest, gaugino masses depend on three parameters, i.e.,$m_{3/2}$,
$|L|$, and $\phi_L$.  Even though there are three free
parameters for three gaugino masses, there exists a non-trivial
constraint as shown in eq.~\eqref{PGMrelation}.  Thus, by checking if
the observed gaugino masses are in agreement with the PGM relation, we
can test the PGM model as a model of SUSY breaking.  Simultaneously,
we can determine the fundamental parameters $m_{3/2}$, $|L|$, and
$\phi_L$.  In particular, with the determination of $|L|$, information
about the masses of Higgsinos and heavy Higgses are obtained (see eq~
\eqref{eq:L_def}).  Furthermore, understanding of the gravitino mass
$m_{3/2}$ will shed light on the mechanism of SUSY breaking.
Determination of the gravitino mass has implications also for
cosmology.  For example, once the gravitino mass is known, an upper
bound on the reheating temperature after inflation can be precisely
determined in order not to spoil the success of big-bang
nucleosynthesis.  For the understanding of the thermal history of the
universe, such an upper bound gives very important information.

In summary, we have demonstrated that the FCC can play crucial roles
in the understanding of physics beyond the standard model.  We have
considered only the PGM model of SUSY breaking, and the importance of
the FCC will become clearer with the studies of other cases.

\acknowledgments

This work was supported by MEXT KAKENHI Grant number JP16K21730 and JSPS KAKENHI Grant (Nos.\ 17J00813 [SC],
16H06490 [TM], 18K03608 [TM], and 18J11405 [MS]).

\end{document}